\documentclass[a4paper,12pt]{article}
\usepackage[dvips]{graphicx}
\usepackage{amsmath}
\title{(3+1)-dimensional relativistic hydrodynamical expansion of hot
and dense matter \\ in ultra-relativistic nuclear collision}

\author{Chiho Nonaka$^1$\thanks{E-mail:nonaka@butsuri.sci.hiroshima-u.ac.jp},  
Eiji Honda$^1$, and Shin Muroya$^2$}

\date{July 18, 2000}

\begin{document}
\maketitle
\begin{center}
\noindent
{\it $^1$Department of Physics, Hiroshima University, \\
Higashi-Hiroshima 739-8526, Japan, and  \\
$^2$Tokuyama Women's College, Tokuyama, 745-8511, Japan
}
\end{center}

\abstract{A full (3+1)-dimensional calculation using the Lagrangian
hydrodynamics is proposed for relativistic nuclear collisions.
The calculation enables us to evaluate anisotropic
flow of hadronic matter which appears in 
non-central and/or asymmetrical relativistic nuclear collisions.
Applying hydrodynamical calculations to the deformed uranium 
collisions at AGS energy region, we discuss the nature of space-time 
structure and particle distributions in detail.
}

\vspace{1cm}



\section{Introduction}
The study of the hot and dense matter which is produced in  
relativistic nuclear collisions has received the intensive attention, 
and a number of experiments 
have been done to realize the phenomena \cite{QM99}.
Hydrodynamical model is one of the established models for 
describing global features in the relativistic nuclear
collisions as collective flow.
Recently, anisotropic flow phenomena have been observed at AGS
\cite{E877f, E895} and also at SPS \cite{NA49, NA49-2, WA98}.
Several authors have argued the relation between 
the behavior of the collective flow and the equation of states. 
Based on the relativistic hydrodynamical model, 
Rischke reported that the existence of the minimum point in the 
excitation 
function of directed flow would suggest the phase transition \cite{Rischke}.
Danielewicz has shown that the elliptic flow is
sensitive to the difference of the nuclear equation of state 
by using a relativistic 
hadron transport model \cite{Danielewicz}.
Sorge has discussed centrality dependence of elliptic flow
based on the event generator RQMD which includes phase transition
\cite{NA49-2,Sorge}. 
Hence, analysis of the flow can inform us the property of 
the nuclear equation of state and quark-gluon plasma (QGP).
 
Since Bjorken proposed the scaling solution \cite{Bjorken}, 
a number of investigations
based on the relativistic hydrodynamical
model have been done.
Those reproduced successfully experimental 
inclusive spectra at both AGS and SPS energies
\cite{Akase}--\cite{Schlei}.
Ornik et al.\  investigated single particle distribution and 
Bose-Einstein correlation at SPS \cite{Ornik}. 
Sollfrank et al.\  discussed the hadron spectra 
and electromagnetic spectra at SPS \cite{Sollfrank}. 
Hung and Shuryak mentioned the equation of states, radial flow, 
and freeze-out at AGS and SPS \cite{Hung}. 
Morita et al.\  discussed single particle distribution and 
Bose-Einstein correlation at SPS \cite{Morita}.
Therefore the relativistic hydrodynamical model works well in the 
analyses of many kinds of phenomena in the ultra-relativistic collision. 
However, in most of the studies based on the relativistic hydrodynamical 
model, cylindrical symmetry is assumed and, therefore, discussions are limited 
only to the central collisions. 

Recently anisotropic flow has been analyzed by using relativistic 
hydrodynamical model. Teaney and Shuryak   
proposed ``nutcracker'' 
phenomenon \cite{Teaney} and Kolb et al.\ discussed anisotropic 
flow and phase transition \cite{Kolb}.   
But these analyses use Bjorken's scaling solution in the longitudinal 
direction, and their discussion is restricted in the mid-rapidity 
region. 

In order to investigate anisotropic flow,  
the quantitatively reliable (3+1)-dimensional relativistic 
hydrodynamical calculation is indispensable.
The full (3+1)-dimensional calculation of relativistic 
hydrodynamical equation has already been done  
by Rischke et al.\ \cite{Rischke2} and Brachmann et al.\
\cite{Brachmann}.
Rischke pointed out that the minimum of the excitation function of directed 
flow suggests the existence of the phase transition \cite{Rischke2}. 
Brachmann et al.\ \cite{Brachmann} discussed the
antiflow of nucleons at the softest point of the equation of 
state using three-fluid dynamics.  
Their numerical schemes are based on the Eulerian hydrodynamics.

Here, we present the Lagrangian hydrodynamic simulation without 
assuming cylindrical
symmetry which makes the full (3+1)-dimensional analyses possible.
The Lagrangian hydrodynamics has several advantages over the 
Eulerian hydrodynamics 
to investigate phenomenon in ultra-relativistic nuclear collisions. 
As is well known, in Lagrangian method, grid points move along the flow, 
and is superior to Eulerian when calculation region varies rapidly.  
This is a great advantage for high energy collision where initial nuclei 
are Lorentz-contracted. 
Secondly, we can follow a trajectory directly in the phase diagram 
in case of Lagrangian algorithm. 
This allows us to study heavy ion collision phenomena together with the 
equation of state. 

As an example of the anisotropic flow in heavy ion collisions, 
we apply our hydrodynamical model to 
deformed uranium-uranium collisions in AGS energy 
and analyze the elliptic flow in detail. 
Shuryak pointed out the remarkable features of the 
deformed uranium collisions which are suitable for the important 
problems such as hard process, elliptic flow, 
and $J/\psi$ suppression \cite{Shuryak}. 
The effect of such deformation to the  flow \cite{B-A.Li} and $J/\Psi$
suppression \cite{Sa} has  been investigated.
Here, we discuss the influence of deformation of uranium nucleus 
on elliptic flow.

In Sec.~II we introduce relativistic hydrodynamical equation and 
explain our original algorithm of the numerical calculation of 
(3+1)-dimensional relativistic hydrodynamical equation.
In Sec.~III we apply our hydrodynamical model to investigation of 
elliptic flow which is produced in deformed uranium collisions and discuss 
the effect of the deformation on elliptic flow in detail.
Section IV is devoted to the summary of the paper.

\section{Algorithm to Solve Hydrodynamical Equation}
\subsection{Relativistic Hydrodynamical Equation}
The relativistic hydrodynamical equation in Lorentz-covariant
form is
\begin{equation}
\partial _{\mu}T^{\mu \nu} = 0.
\label{hydro}
\end{equation}
Since we discuss a system formed by nuclear collision, baryon number
current conservation should be also taken into account,
\begin{equation}
\partial_\nu j^\nu_{B} = 0.
\label{baryon0}
\end{equation}
In this paper,
$T^{\mu \nu}$ is taken as energy-momentum tensor of the perfect fluid,
\begin{equation}
T^{\mu \nu} = \epsilon u^{\mu}u^{\nu}-p(g^{\mu \nu}-
u^{\mu}u^{\nu}),
\end{equation}
and baryon number current is given by
\begin{equation}
j^\nu_{B} = n_B(T,\mu)u^{\nu}.
\label{baryon}
\end{equation}
Here $\epsilon$, $p$ and $n_{B}$
are energy density, pressure and baryon number density, respectively. 
These are the functions of the coordinates through the temperature $T(x_\mu)$
and the baryon number chemical potential $\mu(x_\mu)$. 
$u^\mu = \gamma(1,v_x,v_y,v_z)$
and $g^{\mu \nu}={\rm diag.}(1,-1,-1,-1)$ are local four-velocity
and metric tensor, respectively.
If the equation of state is properly given, we can solve the 
coupled eqs.~(\ref{hydro}) and (\ref{baryon0}) and obtain the
chronological 
evolution of temperature and chemical potential.
In order to make our numerical method clear, 
eqs.~(\ref{hydro}) and (\ref{baryon0}) are rewritten as
\begin{eqnarray}
&
\left (
\begin{array}{ccccc}
\gamma ^{2} v_{x} & \gamma^{2} v_{y} & \gamma ^{2} v_{z} &
\frac{1} {\omega} \frac{\partial \epsilon} {\partial T} &
\frac{1} {\omega} \frac{\partial \epsilon} {\partial \mu} \\
\gamma^{2} & 0 & 0 &
\frac{1} {\omega} v_{x} \frac{\partial p} {\partial T} &
\frac{1} {\omega} v_{x} \frac{\partial p} {\partial \mu} \\
0 & \gamma^{2} & 0 &
\frac{1} {\omega} v_{y} \frac{\partial p} {\partial T} &
\frac{1} {\omega} v_{y} \frac{\partial p} {\partial \mu} \\
0 &0 & \gamma^{2} &
\frac{1} {\omega} v_{z} \frac{\partial p} {\partial T} &
\frac{1} {\omega} v_{z} \frac{\partial p} {\partial \mu} \\
n_{B}\gamma^{2}v_{x} & n_{B}\gamma^{2}v_{y} &
n_{B} \gamma^{2}v_{z} &
\frac{\partial n_{B}} {\partial T} &
\frac{\partial n_{B}} {\partial \mu} \\
\end{array}
\right )
\partial _{t}
\left (
\begin{array}{c}
v_{x} \\ v_{y} \\ v_{z} \\ T \\ \mu
\end{array}
\right )
\nonumber  \\
 + &
\left (
\begin{array}{ccccc}
\gamma^{2}v_{x} ^{2} + 1 & \gamma^{2}v_{x} v_{y} &
\gamma^{2} v_{x}v_{z} &
\frac{1} {\omega} v_{x} \frac{\partial \epsilon} {\partial T} &
\frac{1} {\omega} v_{x} \frac{\partial \epsilon} {\partial \mu} \\
\gamma^{2} v_{x} & 0 & 0 &
\frac{1}{\omega} \frac{\partial p}{\partial T} &
\frac{1}{\omega} \frac{\partial p}{\partial \mu} \\
0 & \gamma^{2} v_{x} & 0 & 0& 0 \\
0 & 0 & \gamma^{2} v_{x} & 0 & 0 \\
n_{B}(\gamma^{2} v_{x}^{2} + 1) &
n_{B} \gamma^{2}v_{x}v_{y} &
n_{B} \gamma^{2}v_{x}v_{z} &
n_{B} \frac{\partial n_{B}}{\partial T} &
n_{B} \frac{\partial n_{B}}{\partial \mu}
\end{array}
\right )
\partial _{x}
\left (
\begin{array}{c}
v_{x} \\ v_{y} \\ v_{z} \\ T \\ \mu
\end{array}
\right )
\nonumber \\
 +&
\left (
\begin{array}{ccccc}
\gamma^{2}v_{y}v_{x} & \gamma^{2} v_{y}^2 +1 &
\gamma^{2} v_{y}v_{z} &
\frac{1}{\omega} v_{y} \frac{\partial \epsilon}{\partial T} &
\frac{1}{\omega} v_{y} \frac{\partial \epsilon}{\partial \mu} \\
\gamma^{2}v_{y} & 0 & 0& 0& 0 \\
0 & \gamma^{2} v_{y}^{2} & 0 &
\frac{1} {\omega} \frac{\partial p} {\partial T} &
\frac{1} {\omega} \frac{\partial p} {\partial \mu} \\
0& 0 & \gamma^{2} v_{y} 0 & 0 \\
n_{B} \gamma^{2}v_{y}v_{x} &
n_{B}(\gamma^{2}v_{y}^{2}+1) & n_{B}\gamma^{2} v_{y} v_{x} &
v_{y}\frac{\partial n_{B}}{\partial T} &
v_{y} \frac{\partial n_{B}}{\partial \mu} \\
\end{array}
\right )
\partial _{y}
\left (
\begin{array}{c}
v_{x} \\ v_{y} \\ v_{z} \\ T \\ \mu
\end{array}
\right )
\nonumber \\
 + &
\left (
\begin{array}{ccccc}
\gamma ^{2} v_{z} v_{x} & \gamma ^{2} v_{z} v_{y} &
\gamma ^{2} v_{z}^{2}+1 &
\frac{1} {\omega} v_{z} \frac{\partial \epsilon}{\partial T} &
\frac{1} {\omega} v_{z} \frac{\partial \epsilon}{\partial \mu} \\
\gamma^{2}v_{z} & 0 & 0 & 0 & 0 \\
0 & \gamma^{2}v_{z} & 0 & 0 & 0\\
0 & 0& \gamma^{2}v_{z} &
\frac{1}{\omega}\frac {\partial p}{\partial T} &
\frac{1}{\omega}\frac {\partial p}{\partial \mu} \\
n_{B}\gamma^{2}v_{z} v_{x} & n_{B}\gamma^{2}v_{z}v_{y} &
n_{B}(\gamma^{2}v_{z}^{2}+1) &
v_{z}\frac{\partial n_{B}}{\partial T} &
v_{z} \frac{\partial n_{B}}{\partial \mu}
\end{array}
\right )
\partial _{z}
\left (
\begin{array}{c}
v_{x} \\ v_{y} \\ v_{z} \\ T \\ \mu
\end{array}
\right )
\nonumber \\
 =  0 ,
\label{nhydro}
\end{eqnarray}
where $\gamma = 1/\sqrt{1-v^2}$, $\omega = \epsilon + p.$
From time-like projection of eq.~(\ref{hydro}),
$u^\nu \partial _{\mu} T^{\mu \nu} = 0 $, and eq.~(\ref{baryon0}),
one can obtain entropy conservation law, 
\begin{equation}
\partial ^ \mu s_\mu = 0,
\label{entropy}
\end{equation}
with the aid of the thermodynamical relation,
\begin{equation}
\epsilon + p=Ts+ \mu n_{B},
\label{thermal}
\end{equation}
where $s^{\mu}$=$su^{\mu}$ is the entropy current density.
 We solve numerically eqs.~(\ref{baryon0}) and (\ref{entropy}) with the
algorithm which will be explained in the next subsection.

\subsection{Computational Scheme}
Most hydrodynamic calculations which are used for investigating
various phenomena in heavy ion collisions are based on the Eulerian 
hydrodynamics.
Sollfrank et al.\ analyze hadron and electromagnetic spectra
by using SHASTA algorithm \cite{Sollfrank}.
HYLANDER and HYLANDER-C algorithm are used by Ornik et al.\ \cite{Ornik}
and Schlei and Strottman \cite{Schlei}, respectively.
Rischke et al.\ use RHHLE algorithm and study hydrodynamics and
collective flow \cite{Rischke}.

Here, we solve the (3+1)-dimensional relativistic hydrodynamical
equation with the Lagrangian hydrodynamics.
The Lagrangian hydrodynamics has several advantages over the Eulerian
hydrodynamics to treat ultra-relativistic nuclear collisions.
At high energies, initial distribution of energy
localizes due to collision of Lorentz contracted projectile and
target.  To treat the situation, fine resolution is required 
in the Eulerian hydrodynamics and computational cost becomes expensive.
On the other hand, in the Lagrangian hydrodynamics,
discretized grids move along the expansion of the fluid, therefore 
we can perform the calculation at all stages on 
the lattice points which we prepare in the initial condition. 
For example, in our previous calculation \cite{Nonaka} 
the fluid expands four times larger in longitudinal
direction. This fact means we need four times number of
grid if we use naive Eulerian type algorithm. 
Another merit of the Lagrangian hydrodynamics is that it enables us to
derive the physical information directly, because it follows
the flux of the current. 
For example, the path of a volume element of fluid in the $T$-$\mu$
plane can be traced as we will demonstrate in the next subsection.
Therefore, we are able to discuss how the phase between hadron phase 
and QGP phase effects the physical phenomena by 
 the Lagrangian hydrodynamics.

Our numerical calculation of the (3+1)-dimensional
relativistic hydrodynamical equation is as follows:
First, the coordinate
at time $t+\Delta t$, 
$x^{m} = X^{m}(t,i,j,k)$ is evolved as,
\begin{equation}
X^{m}(t+\Delta t,i,j,k)=X^{m}(t,i,j,k) + \frac{u^{m}(t,i,j,k)}
{u^{t}(t,i,j,k)}\Delta t.
\label{cood}
\end{equation}
By definition of the Lagrangian hydrodynamics, the coordinates move in parallel
with $j^\mu$ and $s^\mu$.

Second, the local velocity is determined, 
\begin{eqnarray}
v^{m}(t+\Delta t,i,j,k) & = &  v^{m}(t,i,j,k) + 
\partial_{t} v^{t}(i,j,k,t) \Delta t
\nonumber \\
& + & \sum _{n=1} ^{3} \partial_{n}v^{m}(i,j,k,t)
(X^{n}(t+\Delta t,i,j,k) - X^{n}(t,i,j,k)),
\end{eqnarray}
where the value of $\partial ^{\mu}v_{\mu}$ is obtained 
from eq.~(\ref{nhydro}).

Finally, the temperature and chemical potential are derived. 
The volume element $d\sigma^{\mu}$  at time $t$ is surrounded by 
 eight points, $X^\mu(t,i,j,k)$, $X^\mu(t,i+1,j,k)$, 
$X^\mu(t,i,j+1,k)$, 
$X^\mu(t,i,j,k+1)$, $X^\mu(t,i+1,j+1,k)$, $\cdots$, 
$X^\mu(t,i+1,j+1,k+1)$. 
Using this volume element,   
eqs.~(\ref{baryon0}) and (\ref{entropy}) are 
rewritten as   
\begin{eqnarray}
\lefteqn{s(T(t+ \Delta t,i,j,k),\mu(t+\Delta t, i,j,k))
u^{t}(t+ \Delta t, i,j,k) d \sigma_{t}(t+\Delta t,i,j,k)}
\hspace{3cm}
\nonumber \\ 
&=&   s(T(t,i,j,k),\mu(t, i,j,k))
u^{t}(t, i,j,k) d \sigma_{t}(t,i,j,k),
\label{eqn:dt}
\end{eqnarray}
\begin{eqnarray}
\lefteqn{n_{B}(T(t+\Delta t,i,j,k),\mu(t+\Delta t, i,j,k))
u^{t}(t+\Delta t, i,j,k) d \sigma_{t}(t+\Delta t,i,j,k)}
\hspace{3cm}
\nonumber \\
&= &n_{B}(T(t,i,j,k),\mu(t, i,j,k))
u^{t}(t, i,j,k)d \sigma_{t}(t,i,j,k).
\label{eqn:dnb}
\end{eqnarray}
Here, by virtue of the determination of coordinates eq.~(\ref{cood}), 
we can use the relation,  
\[
u^\mu d\sigma _\mu = u^t d\sigma_t.
\]
Since $s$ and $n_B$ depend on  $T$ and $\mu$,   
using up to the first order of the differences of temperature,    
$
\Delta T (t,i,j,k) \equiv T(t+\Delta t,i,j,k)-T(t,i,j,k) 
$, and of the chemical potential,  
$
\Delta \mu (t,i,j,k) \equiv \mu(t+\Delta t,i,j,k)-\mu(t,i,j,k) 
$, 
we expand  $s$ and $n_B$  as, 
\begin{eqnarray}
\lefteqn{
s(T(t+\Delta t,i,j,k),\mu(t+\Delta t,i,j,k))}\hspace{1cm}
\nonumber \\
&=& s(T(t,i,j,k),\mu(t,i,j,k)) \nonumber \\ 
& & { } + \left[ 
\frac{\partial s}{\partial T}
\right ]_{\footnotesize
\begin{array}{c}
 T=T(t,i,j,k) \\
\mu = \mu (t,i,j,k) 
\end{array}
} \Delta T+ 
\left [
\frac{\partial s}{\partial \mu}
\right]_{\footnotesize
\begin{array}{c}
 T=T(t,i,j,k) \\
\mu = \mu (t,i,j,k) 
\end{array}
}\Delta \mu,
\label{dels}
\end{eqnarray}
\begin{eqnarray}
\lefteqn{n_B(T(t+\Delta t,i,j,k),\mu(t+\Delta t,i,j,k))} \hspace{1cm}
\nonumber \\
&= &n_B(T(t,i,j,k),\mu(t,i,j,k)) \nonumber \\
& & { } +
\left[ 
\frac{\partial n_B}{\partial T}
\right ]_{\footnotesize
\begin{array}{c}
 T=T(t,i,j,k) \\
\mu = \mu (t,i,j,k) 
\end{array}
} \Delta T+ 
\left [
\frac{\partial n_B}{\partial \mu}
\right]_{\footnotesize
\begin{array}{c}
 T=T(t,i,j,k) \\
\mu = \mu (t,i,j,k) 
\end{array}
}\Delta \mu .
\label{delnb}
\end{eqnarray}
Substituting eqs.~(\ref{dels}) and (\ref{delnb}) 
to eqs.~(\ref{eqn:dt}) and (\ref{eqn:dnb}), we obtain the temperature 
and chemical potential at the next time step,
\begin{eqnarray}
\lefteqn{
T(t+\Delta t,i,j,k)= T(t,i,j,k)}\hspace{12cm} 
\nonumber \\
+ \frac{1}{\Delta _{s,n_{B}}} 
\left . \left \{
\frac{\partial n_{B}}{\partial \mu} s(T,\mu)
- \frac{\partial s}{\partial \mu} n_{B}(T,\mu)
\right \}  \right |_{\footnotesize
\begin{array}{c}
 T=T(t,i,j,k) \\
\mu = \mu (t,i,j,k) 
\end{array}
}
\left [\cdots 
\right ],
\end{eqnarray}
\begin{eqnarray}
\lefteqn{
\mu(t+\Delta t,i,j,k) = \mu(t,i,j,k)}\hspace{12cm} 
\nonumber \\
+ \frac{1}{\Delta _{s,n_{B}}} 
\left .\left \{
\frac{\partial n_{B}}{\partial T} n_{B}(T,\mu)
- \frac{\partial s}{\partial T} s(T,\mu)
\right \}  \right |_{\footnotesize
\begin{array}{c}
 T=T(t,i,j,k) \\
\mu = \mu (t,i,j,k) 
\end{array}
}
\left [\cdots
\right ],
\end{eqnarray}
where $\Delta_{s,n_B}$ and $[\cdots]$ are 
\[
\Delta_{s,n_{B}} = \left . \left (\frac{\partial s(T,\mu)}{\partial T} 
\frac{\partial n_{B}(T,\mu)} {\partial \mu} -
\frac{\partial s(T,\mu)}{\partial \mu}
\frac{\partial n_{B}(T,\mu)}{\partial T}
\right ) \right |_{\footnotesize
\begin{array}{c}
 T=T(t,i,j,k) \\
\mu = \mu (t,i,j,k) 
\end{array}
}, 
\]
\[ \displaystyle
\left [ \cdots \right ] = 
\left [ 
\frac{u^{t}(t,i,j,k)d \sigma ^{t}(t,i,j,k)}
{u^{t}(t+\Delta t,i,j,k)d\sigma^{t}(t+\Delta t,i,j,k)} -1
\right ] ,
\]
respectively. These numerical procedures are the extension of the 
method in ref.\ 
\cite{Ishii}.

In this algorithm CPU time is almost proportional to the number of 
the lattice point.
Numerical calculation of the relativistic hydrodynamical 
equation in this paper has been performed 
 at the Institute
for Nonlinear Sciences and Applied Mathematics, Hiroshima University.
Average floating point operations for $(55,63,49)$ 
space points and 5300 time steps are 36 Tera in 
each calculation reported in the next section.

\begin{figure}[h]
\begin{center}
\begin{minipage}{0.45 \linewidth}
\includegraphics[width=\linewidth]{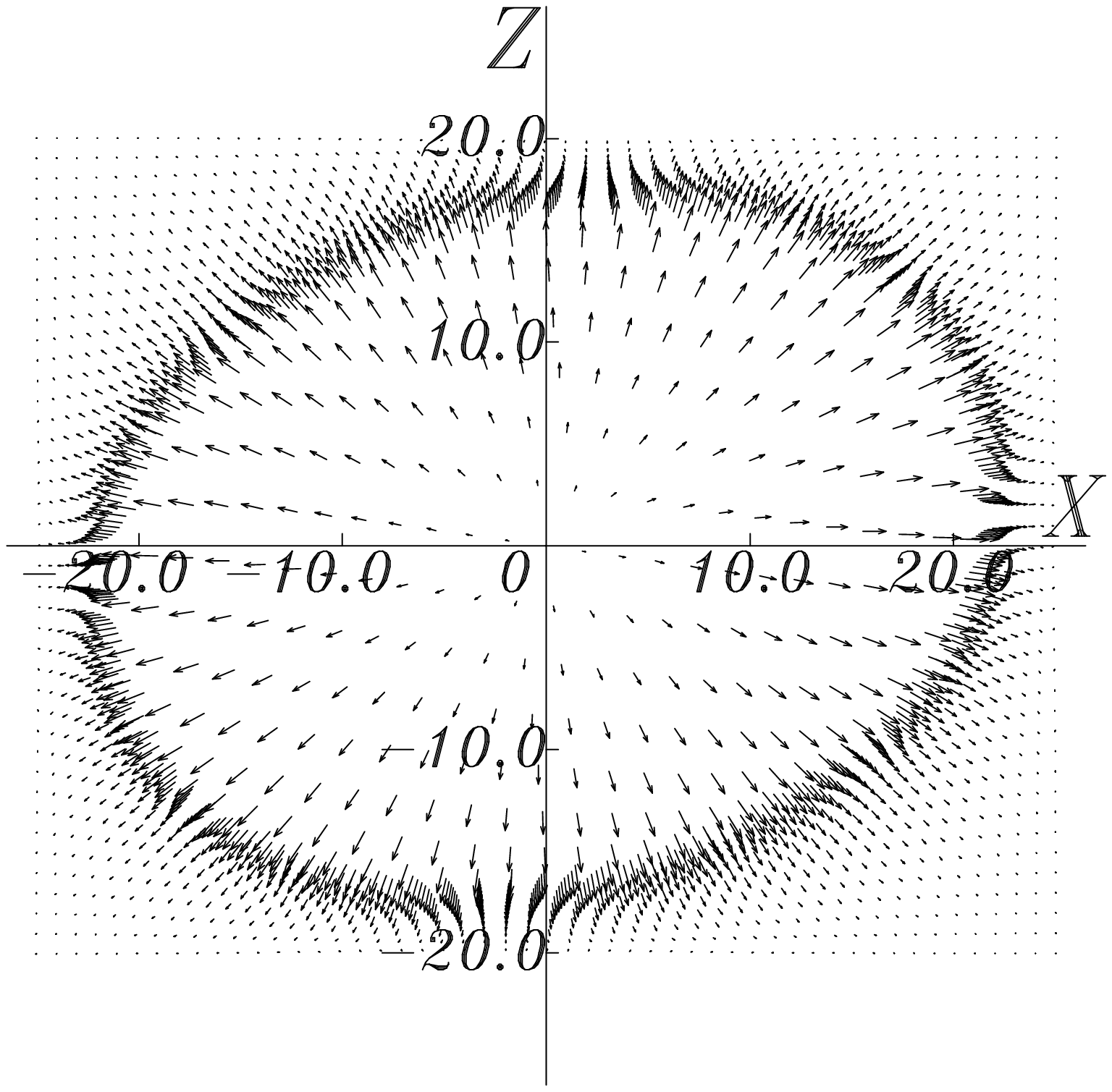}
\end{minipage}
\hspace{0.4cm}
\begin{minipage}{0.45\linewidth}
\includegraphics[width=\linewidth]{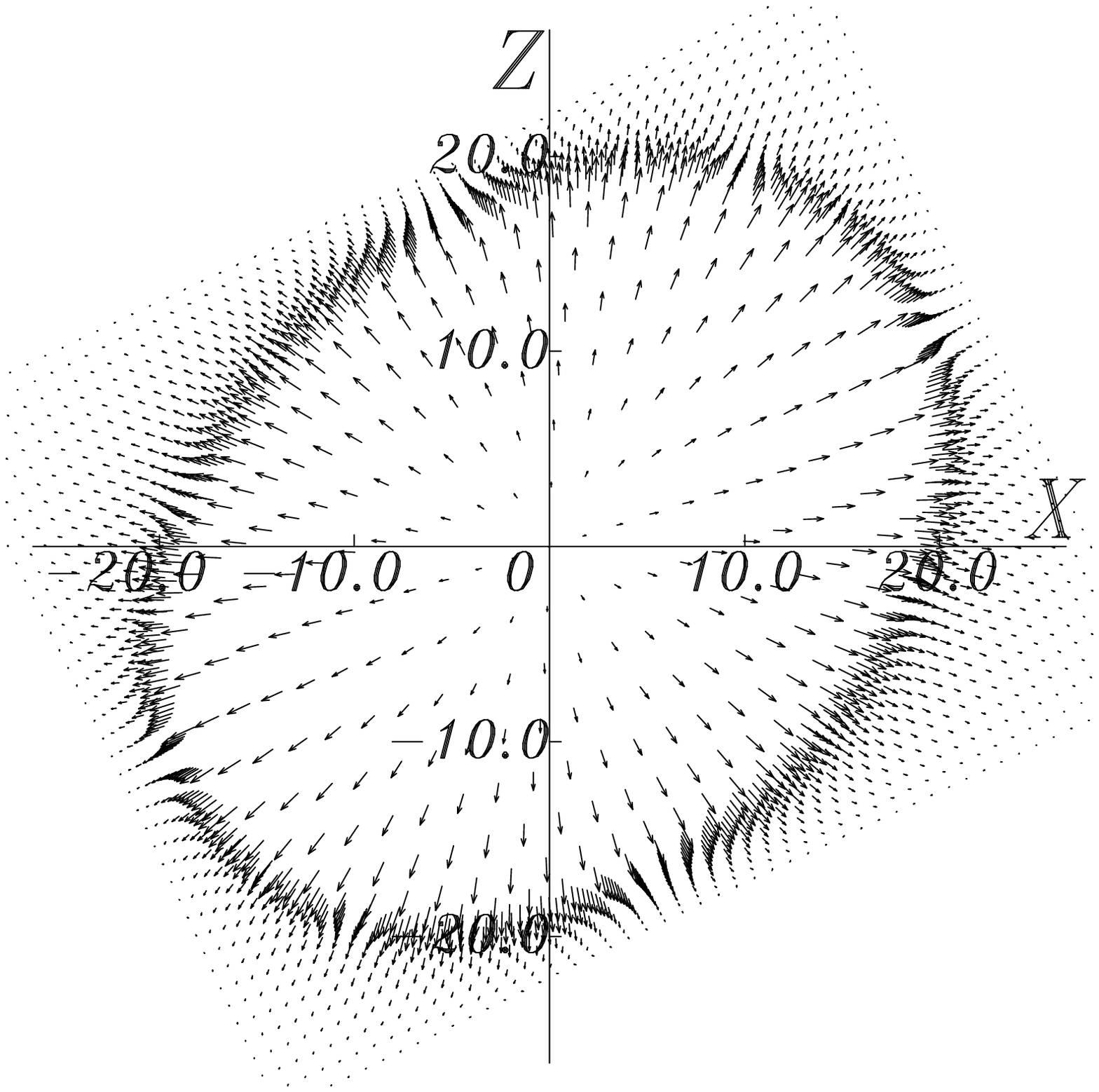}
\end{minipage}
\caption{The left part of the figure indicates the flow at $y \sim 0$ fm 
and $t = 20 $ fm/$c$ in Au + Au 10.8 AGeV collision \cite{Nonaka}. 
The right part of the figure shows the same result which is obtained by 
the calculation with the grids rotated 26 degree on the $x$-$z$ plane. }
\label{flow}
\end{center}
\end{figure}

In order to analyze anisotropic flow with high accuracy, 
the artificial anisotropy which can be caused by the discretization of space 
should be small. We checked the reliability of our calculation 
by comparing the results with rotated spatial grid. 
Figure \ref{flow} shows that results of flow obtained by  
different choice of the grid.  
The difference of flow   
between two results is
less than  0.15 \% in the present calculations. 

\subsection{Path in the Phase Diagram}
\begin{figure}[h]
\begin{center}
\begin{minipage}{0.4 \linewidth}
\includegraphics[width=\linewidth]{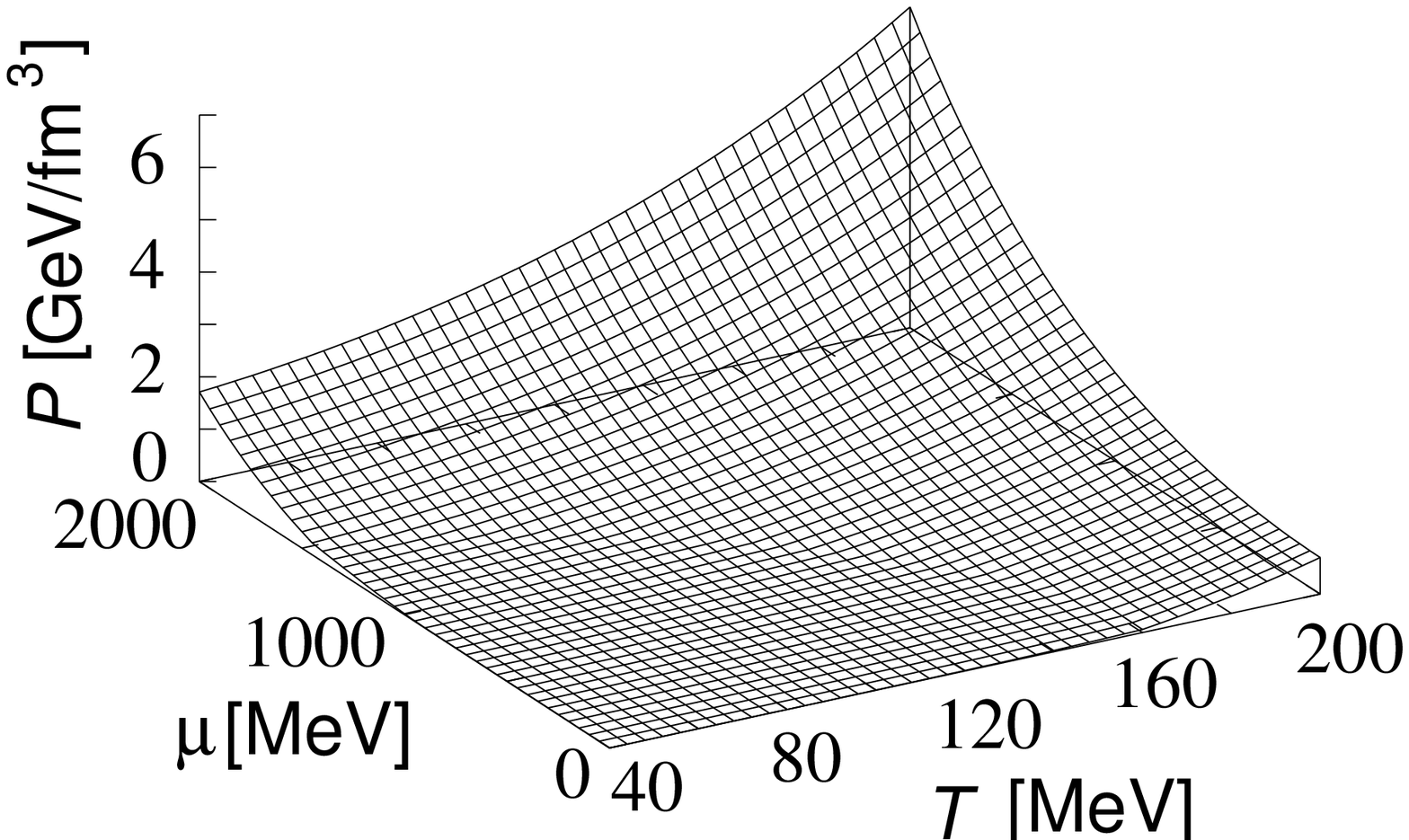}
\caption{The equation of states including the first order phase transition.}
\label{eos}
\end{minipage}
\hspace{0.4cm}
\begin{minipage}{0.4\linewidth}
\includegraphics[width=\linewidth]{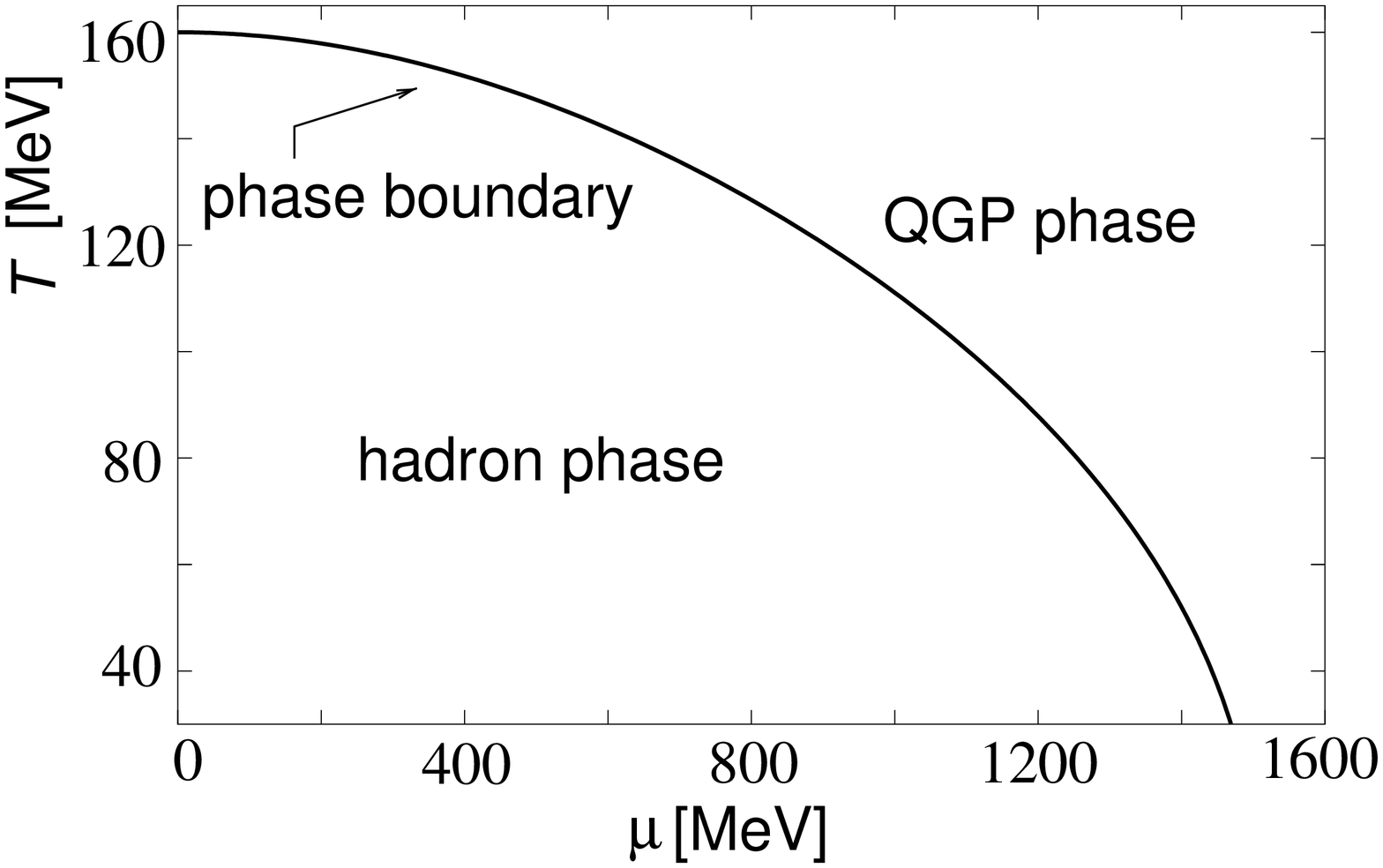}
\caption{The phase boundary which is determined by the pressure balance.}
\label{phase-b}
\end{minipage}
\end{center}
\end{figure}

The Lagrangian hydrodynamics enables us to trace easily 
the history of the trajectory of flux.  
In order to test the applicability of our algorithm to study 
the chronological trajectory of the volume 
element in the phase diagram, we use  
the equation of state which contains the first order phase transition 
only in this subsection. 
Above the phase transition, the thermodynamical quantities are 
assumed to be determined by QGP gas which  
is dominated by massless $u,d,s$ quarks and gluons.
In the QGP phase the pressure is given as, 
\begin{equation}
p = \frac{(32+21N_f)\pi^2}{180}T^4 + \frac{N_f}{2}\left ( \frac{\mu}{3} 
\right )^2 T^2 + \frac{N_f}{4\pi^2}\left( \frac{\mu}{3}\right ) ^4-B,
\end{equation}
where $N_f$ is 3 and $B$ is Bag constant \cite{Sollfrank,Hung}. 
For the hadron phase we use the excluded volume model \cite{Rischke3} 
which contains all resonances up to 2.0 GeV \cite{Particle}.
In the hadron phase the pressure for fermion is given as, 
\begin{eqnarray}
p^{excl}(T,\{ \mu_i \}) & = & \sum_i p_i^{ideal}(T,\mu_i-V_0 p^{excl}
(T,\{ \mu_i\})), \nonumber \\
 & = & \sum _i p_i ^{ideal} (T, \tilde{\mu_i}), 
\end{eqnarray}
where $p^{ideal}$ is the pressure of ideal hadron gas and $V_0$ is
excluded volume of which radius is fixed to 0.7 fm. 
Putting the critical temperature as 160 MeV for zero chemical
potential, 
the Bag constant, $B^\frac{1}{4}$, is given as 233 MeV.
Figure \ref{eos} shows the equation of states as a function of 
temperature and chemical potential.
Figure \ref{phase-b} indicates the phase boundary which is 
determined by the pressure balance between the 
two phases, i.e., $p_Q = p_H$.
In the mixed phase we introduce the fraction of the 
volume of the QGP phase, $\lambda(x_\mu)$ $(0 \leq \lambda \leq 1)$ and 
parameterize energy density and baryon number density as,  
\begin{eqnarray}
\epsilon_M(\lambda,T^*(\mu)) & = & \lambda \epsilon_Q(T^*(\mu))-(1-\lambda)
\epsilon_H(T^*(\mu)),  \nonumber \\
n_{BM}(\lambda,T^*(\mu)) & = & \lambda n_{BQ}(T^*(\mu))-(1-\lambda)
n_{BH}(T^*(\mu)), 
\end{eqnarray}
where  $T^*(\mu)$ is the value of temperature on 
the phase boundary in phase diagram.
Contrary to in the Eulerian hydrodynamics where the boundary condition 
should be considered 
on the discontinuous plane between two phases \cite{Landau, Gyulassy},  
in our algorithm, by virtue of the explicit use of the 
current conservation equations,
the flux of the fluid can be traced easily  
even if the discontinuity of the first order phase transition 
exists \cite{Morita,Nonaka2}.  
For initial conditions, we use the results which are obtained 
by URASiMA (Ultra-Relativistic A-A collision Simulator 
based on Multiple scattering Algorithm) 
in Au+Au 20 AGeV collisions \cite{Nonaka}. 

\begin{figure}[t]
\begin{center}
\begin{minipage}{0.6\linewidth}
\includegraphics[width=\linewidth]{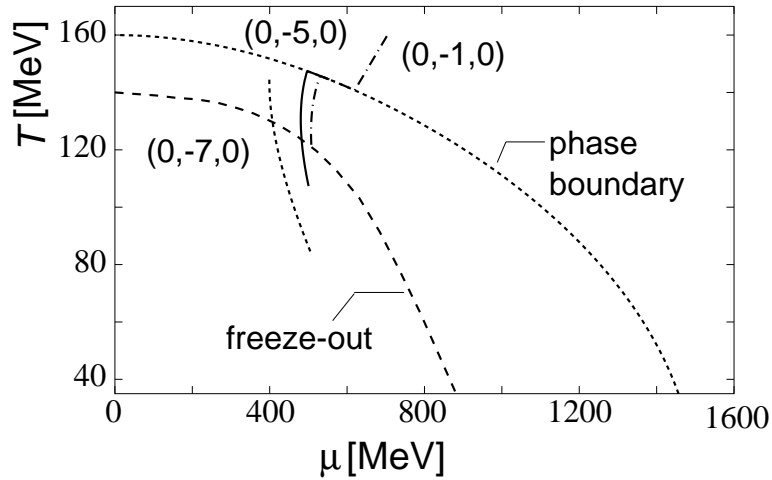}
\end{minipage}
\end{center}
\caption{The paths in the phase diagram. The dot-dashed line stands
 for the path which starts from the QGP phase and on this line 
$n_B/s $ is 0.065. 
The solid line stands for 
 the path which starts from the mixed phase, and on this line 
$n_B/s$ is 0.059. The dotted line stands for 
the path which starts from the hadron phase, and on this line $n_B/s$ 
is 0.044.}
\label{phase-d}
\end{figure}

Figure \ref{phase-d} shows the typical paths in the phase diagram.
For instance, the trajectory of the volume element of the grid number  
$(x,y,z) = (0,-1,0)$, which starts from QGP region, moves along phase 
boundary before entering the hadron phase.
The volume element of $(0,-5,0)$, which starts from the mixed phase,
also moves along phase boundary before turning into the hadron phase.
On the other hand the volume element of the grid number $(x,y,z) =
(0,-7,0)$, which starts from hadron phase, draws smooth trajectory 
to the freeze-out. This behavior is the same result as discussed in ref.~\cite{Hung}.
By virtue of  the Lagrangian hydrodynamics we can easily trace the trajectory 
which corresponds to the adiabatic paths in the $T$-$\mu$ plane.

\section{Effect of Deformed Uranium on Collective Flow} 
As an application of our hydrodynamical model, 
we investigate the flow which is generated in the deformed uranium 
collisions.
Shuryak has pointed out that remarkable problems such as hard processes,
elliptic flow, and the mechanism of $J/\psi$ suppression can 
be resolved by using the deformed uranium collisions \cite{Shuryak}.
Since Danielewicz showed that elliptic flow is 
sensitive on nuclear equation of state in AGS energy \cite{Danielewicz},
the elliptic flow is one of the hottest topics in high energy 
nuclear physics.
The high accuracy experiments for elliptic flow have been done 
at AGS \cite{E877f, E895} and SPS \cite{NA49, NA49-2, WA98}.  
Recently, using the cascade model of the ART, Li discussed 
the elliptic flow in the deformed uranium 
collisions. 
If the deformation causes a large influence on anisotropic flow, 
the analyses of collective flow using U+U collisions are promising 
for investigating the difference between QGP states and hadron states. 
In order to analyze anisotropic flow of deformed uranium collisions, 
the (3+1)-dimensional relativistic hydrodynamical model plays 
a central role, i.e. it provides us reliable quantitative 
results. 

Here, the ellipticity is measured by the asymmetry of azimuthal 
distribution of particle which is expanded  
based on Fourier series,  
\begin{equation}
\frac{dN}{d\phi} \sim v_0 (1+2v_1 \cos (\phi) + 2v_2 \cos (2\phi)),
\label{v2}
\end{equation}
where $\phi$ is azimuth and $v_0$ is normalization.
Parameters $v_1$ and  $v_2$ correspond to the intensity of the directed flow and 
elliptic flow, respectively. 

The shape of deformed uranium nucleus is approximately ellipsoid and  
short ($R_t$) and long ($R_c$) semi-axis are given as, 
\begin{eqnarray}
R_t & = & R_s(1-\frac{1}{3} \delta), \nonumber \\
R_c & = & R_s(1+\frac{2}{3}\delta),
\label{ura}
\end{eqnarray}
where $\delta$ =0.27 is the deformation parameter \cite{Bohr}. 
We will investigate how the deformation and the 
orientation between 
two  colliding deformed uranium nuclei influence the flow of 
produced hadrons.  
Among many types of collisions for the orientation, 
we focus on two types of collision, i.e., tip-tip collision in which 
the long axes of two nuclei are along beam direction, and body-body 
collision in 
which the long axes of two nuclei are parallel each other but 
perpendicular to the beam 
direction. 
We calculate also sphere-sphere collision for comparison.

\subsection{Model Description}
The outline of our calculation procedure is as follows:
First, we parameterize the initial conditions of energy 
density, baryon number density and local velocity based on 
the result of event generator URASiMA. 

Our event generator URASiMA \cite{Date,Sasaki1,Sasaki2} is
characterized by multi-chain model (MCM) by which multi-particle
production process can be described successfully.  In 
URASiMA the detailed balance between quasi-two-body production
and absorption processes holds.
It is applicable to AGS and SPS energy regions and
calculated  results reproduce experimental data of
hadron spectra \cite{Date}. Recently thermodynamical properties
of hot and dense hadronic gas are also investigated by URASiMA
\cite{Sasaki1,Sasaki2}. For a more detailed discussions, see refs.\
\cite{Date,Sasaki1,Sasaki2}.

In order to solve the relativistic hydrodynamical equation, 
we need to introduce an equation of state.
Since our calculation does not rely on any artificial assumption,
we can investigate how
the difference of equation of state has an effect on physical
phenomena.
For the first trial, we adopt the equation of state of the 
ideal hadron gas including resonances; this 
is the same equation of state as we used in the 
initial conditions.
The temperature and chemical potential of volume elements
vary with the space-time evolution of fluid until hadronization
process occurs.

\begin{figure}
\begin{center}
\begin{minipage}{0.5\linewidth}
\includegraphics[width= \linewidth]{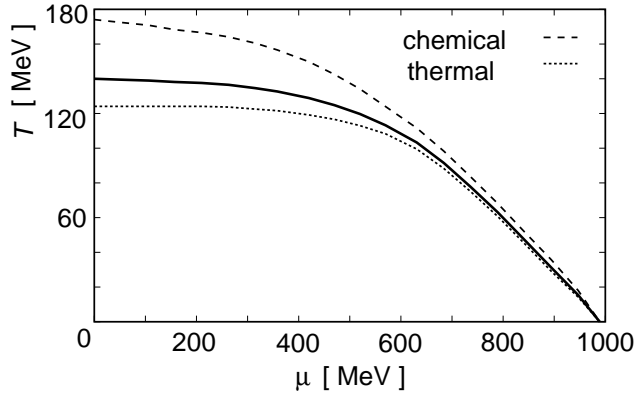}
\caption{The freeze-out condition.
The solid line stands for freeze-out condition which is 
determined from thermal freeze-out (the dotted line) and 
chemical freeze-out (the dashed line).}
\label{fig1}
\end{minipage}
\end{center}
\end{figure}

Finally hadron spectra are obtained by Cooper-Frye formula \cite{Cooper-Frye}.
We assume that the hadronization process occurs when the temperature and
chemical potential of the volume elements cross the boundary (solid
line in fig.~\ref{fig1}).
The solid line is obtained so that the
freeze-out temperature
becomes 140 MeV at vanishing chemical potential, based on chemical
freeze-out model and thermal freeze-out model \cite{freeze-out}.
Using Cooper-Frye formula the particle distribution is given as, 
\begin{equation}
E \frac{dN}{d^3P}=\sum_h \frac{g_h}{(2\pi)^3}
\int_\sigma d\sigma_{\mu}P^{\mu}
\frac{1}{\exp[(P_\nu u^\nu - \mu _f)/T_f] \pm 1},
\label{C-F}
\end{equation}
where $g_h$ is degeneracy of hadrons and $T_f$ and $\mu_f$ are the
freeze-out temperature and chemical potential shown in fig.~\ref{fig1}, 
respectively and $u^\mu$ is local velocity of the fluid on
 the hypersurface $d\sigma_\mu$.
We determine hypersurface $d\sigma_\mu$ by evaluating the normal
 vector to the freeze-out hypersurface  \cite{QGP}.
We evaluate intensity of the elliptic flow $v_2$ 
using eqs.~(\ref{v2}) and (\ref{C-F}).

\subsection{Initial Conditions}
In high energy collisions such as SPS and RHIC, the contribution of 
the spectator to initial conditions is small and ``in-plane'' elliptic 
flow is enhanced \cite{Ollitrault}. 
However, if we focus on AGS energy regions in which 
incident energy is not so large,  
 the effect of deformation can remain strong. 
In order to prepare appropriate initial conditions, 
we estimate the energy density 
distribution and baryon number 
density distribution by the event generator URASiMA.  
We calculate the space-time
 evolution in U+U 10 AGeV and 20 AGeV tip-tip, body-body and 
sphere-sphere collisions at $b=$ 0 and 6 fm,
using the event generator URASiMA, where $b$ is the impact parameter.
In the case of tip-tip and body-body collisions, 
we consider the collision of the deformed uranium according 
to eq.~(\ref{ura}). 
We assume that the hydrodynamical expansion starts when the projectile 
nucleus finishes passing through the target nucleus. 
At this time initial conditions of hydrodynamical model, 
$T(x_\mu)$, $\mu(x_\mu)$, and $u^\mu(x_\mu)$ should be given. 
\begin{table}
\caption{The energy and baryon number densities obtained by URASiMA at 
$(x,y,z)=(0,0,0)$ for different type of collision and the impact 
parameter. (a) U+U 10 AGeV, (b) U+U 20 AGeV}
\begin{center}
Table 1(a): The result of URASiMA (U+U 10.0 AGeV). 
\begin{tabular}{|c|c|c|c|c|}
\hline 
$b$ [fm] & type of collision & 
initial time [fm/$c$]
&
\begin{tabular}{c}
  $\epsilon$ 
 [GeV/fm$^3$] \\   
at $(0,0,0)$ 
\end{tabular}
& 
\begin{tabular}{c}
$n_B$ [fm$^{-3}$] \\
 at $(0,0,0)$
\end{tabular}
\\
\hline
0 & tip-tip & 8.5 & 1.975 & 0.977 \\
\hline
6 & tip-tip & 8.5 & 1.403 & 0.718 \\
\hline
0 & sphere-sphere & 7.0 & 1.807 & 0.929 \\
\hline
6 & sphere-sphere & 7.0 & 1.611 & 0.809 \\
\hline
0 & body-body & 6.5 & 1.681 & 0.819 \\
\hline
6 & body-body & 6.5 & 1.490 & 0.728 \\
\hline
\end{tabular}
\label{uu10-u}
\end{center}
\begin{center}
Table 1(b): The result of URASiMA (U+U 20.0 AGeV).
\begin{tabular}{|c|c|c|c|c|}
\hline 
$b$ [fm] & type of collision &
initial time [fm/$c$]
&
\begin{tabular}{c}
 $\epsilon$  [GeV/fm$^3$] \\
at $(0,0,0)$ 
\end{tabular}
& 
\begin{tabular}{c}
$n_B$ [fm$^{-3}$] \\ at $(0,0,0)$
\end{tabular}
\\
\hline
0 & tip-tip & 6.0 & 2.871 & 1.124 \\
\hline
6 & tip-tip & 6.0 & 2.060 & 0.825 \\
\hline
0 & sphere-sphere & 5.0 & 2.744 & 1.064 \\
\hline
6 & sphere-sphere & 5.0 & 2.241 & 0.869 \\
\hline
0 & body-body & 4.5 & 2.248 & 0.933 \\
\hline
6 & body-body & 4.5 & 2.253 & 0.860 \\
\hline
\end{tabular}
\label{uu20-u}
\end{center}
\end{table}
The results of URASiMA are listed in table \ref{uu10-u}(a) (U+U 10 AGeV) 
and table \ref{uu20-u}(b) (U+U 20 AGeV). 
In body-body collision the energy density and baryon number 
density are the smallest among different types of collision 
in the same incident energy and impact parameter.
For tip-tip collision a decrease in the energy density and 
baryon density to impact parameter is the largest among the three 
types of collision. 
Then the initial energy density distribution and baryon number 
density distribution are parametrized based on these data in 
tables \ref{uu10-u}(a) and \ref{uu20-u}(b). 
The initial energy density and baryon number density distributions are 
given as, 
\begin{eqnarray}
\epsilon &=& \epsilon _{\rm max} B(x,y,z), 
\nonumber \\
n_B & = & n_{B {\rm max}} B(x,y,z), 
\label{initial}
\end{eqnarray}
where $B(x,y,z)$ is the distribution function, which is determined 
so that the result of URASiMA is reproduced, and $\epsilon_{\rm max}$ 
and $n_{B{\rm max}}$ are the values of the result of URASiMA in central 
region.
We interpolate and/or extrapolate the values of $\epsilon_{\rm max}$  
and $n_{B{\rm max}}$ from the data of tables \ref{uu10-u}(a) and
\ref{uu20-u}(b) as follows: 
The distribution function is given as,
\begin{eqnarray}
B(x,y,z)& = & \frac{1}{c_1}[a_1\exp (-\frac{(x-x_{cs})^2}{\sigma _{sx} ^2}
-\frac{(y-y_{cs})^2}{\sigma _{sy} ^2}
-\frac{(z-z_{cs})^2}{\sigma _{sz} ^2}) 
\nonumber \\
& & +a_1\exp (-\frac{(x+x_{cs})^2}{\sigma _{sx} ^2}
-\frac{(y+y_{cs})^2}{\sigma _{sy} ^2}
-\frac{(z+z_{cs})^2}{\sigma _{sz} ^2})
\nonumber \\
& & +\exp (-\frac{(x-x_{cp})^2}{\sigma _{px} ^2}
-\frac{(y-y_{cp})^2}{\sigma _{py} ^2}
-\frac{(z-z_{cp})^2}{\sigma _{pz} ^2})
],
\label{B}
\end{eqnarray}
where $c_1$ is normalization and the parameter,  
which corresponds to the ratio of energy density 
 of spectator to that of participant, $a_1$ is fixed to be 
0.7 for all cases. The ratio of baryon number density is also 
fixed by $a_1$. 
In eq.~(\ref{B}) $x_{cp}$, $y_{cp}$, and $z_{cp}$ are the
center of participant, and $x_{cs}$, $y_{cs}$ and $z_{cs}$ are the center of 
spectator.
Their specific values are determined geometrically by the position of 
projectile nucleus and target nucleus.
For the participant,  
\begin{eqnarray}
x_{cp}& = & y_{cp} = z_{cp}=0,  
\nonumber
\\
\sigma _ {px} & = & a_3(R_x-\frac{b}{2}),
\nonumber \\
\sigma _ {py} & = & \frac{R_y}{R_x}\sqrt{R_x ^2-\frac{b^2}{4}},
\nonumber \\
\sigma_{pz} & = & a_4\frac{z_{cs}}{\gamma}.
\label{part}
\end{eqnarray}
For the spectator, 
\begin{eqnarray}
x_{cs} & = & \frac{R_x}{2}+\frac{b}{4}, 
\nonumber \\
y_{cs} & = & 0,
\nonumber \\
z_{cs} & = & t_i - \frac{R_z}{\gamma},
\nonumber \\
\sigma_{sz} & = & a_5\frac{b}{2},
\nonumber \\
\sigma_{sy} & = & 
\left \{
\begin{array}{lr}
R_x &
(b \geq R_x)
\\
\frac{R_y}{R_x}\sqrt{2R_xb-b^2}
&(b \leq R_x)
\end{array}
\right .,
\nonumber \\
\sigma_{sz} & = & a_6\frac{R_z}{R_y \gamma}.
\label{spec}
\end{eqnarray} 
$R_x,R_y$, and $R_z$ are the radius of projectile nucleus and 
target nucleus in 
the $x$, $y$, and $z$ directions.
In eq.~(\ref{spec}) 
$t_i$ is the initial time when the hydrodynamical evolution starts.
In eqs.~(\ref{part}) and (\ref{spec}), parameters $a_3 \sim a_6$ are 
tuned so that the energy density and baryon number density distribution, 
eq.~(\ref{initial}), reproduce the result of URASiMA. 
The parameters $a_3 \sim a_6$ 
are fixed to 0.7, 2.5, 1.3, 0.7, respectively.
\begin{figure}
\begin{center}
\includegraphics[width= 0.5\linewidth]{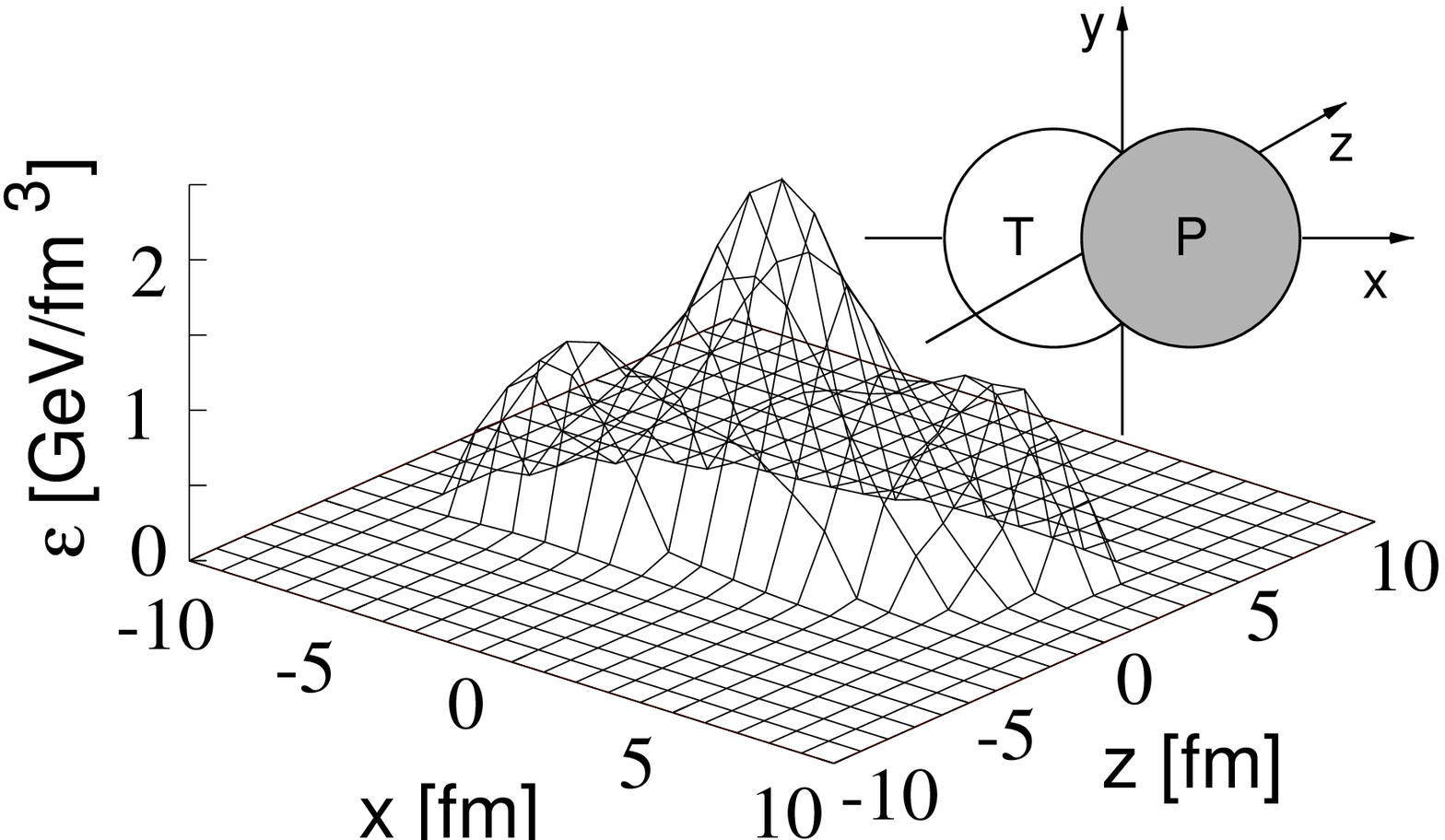}
\caption{The initial condition of energy density distribution 
at $y=0$ fm in U+U 20 AGeV, sphere-sphere collision. 
In this case the impact parameter is 4.0 fm.}
\label{fig2}
\end{center}

\begin{center}
\includegraphics[width= 0.5\linewidth]{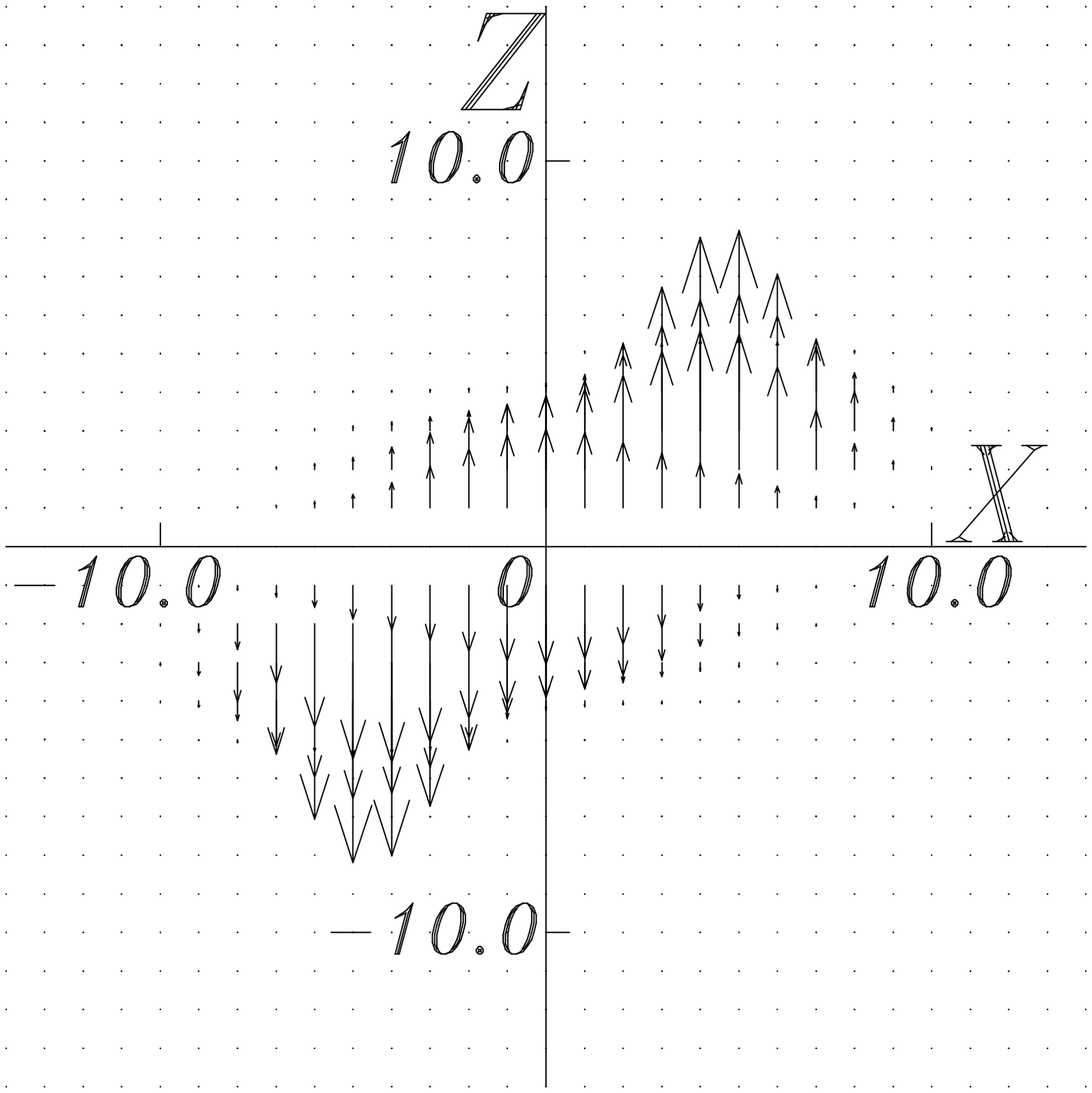}
\caption{The initial condition of velocity distribution 
at $y=0$ fm in the same condition as fig.~\ref{fig2}. }
\label{fig3}
\end{center}
\end{figure}
Figure \ref{fig2} shows the energy density distribution of initial
conditions in U+U 20 AGeV sphere-sphere collision at $b=4.0$ fm.
From this figure we can see clearly the contribution from participant part 
and spectator part.  
For the initial velocity distribution we neglect the transverse flow 
because its value is small in the result of URASiMA. 
The initial flow distribution is given as,
\begin{eqnarray}
v_x(x,y,z)& = & 0,
\nonumber \\
v_y(x,y,z) & = & 0,
\nonumber \\
v_z(x,y,z) & = & \frac{z}{t}B_v(x,y,z).
\end{eqnarray}
Here the distribution function $B_v(x,y,z)$ is given by,
\begin{eqnarray}
B_v(x,y,z)& = & \frac{1}{c_2}[\exp (-\frac{(x-x_{cs})^2}{\sigma _{sx} ^2}
-\frac{(y-y_{cs})^2}{\sigma _{sy} ^2}
-\frac{(z-z_{cs})^2}{\sigma _{sz} ^2}) 
\nonumber \\
& & +\exp (-\frac{(x+x_{cs})^2}{\sigma _{sx} ^2}
-\frac{(y+y_{cs})^2}{\sigma _{sy} ^2}
-\frac{(z+z_{cs})^2}{\sigma _{sz} ^2})
\nonumber \\
& & +a_2\exp (-\frac{(x-x_{cp})^2}{\sigma _{px} ^2}
-\frac{(y-y_{cp})^2}{\sigma _{py} ^2}
-\frac{(z-z_{cp})^2}{\sigma _{pz} ^2})
],
\end{eqnarray}
where $c_2$ is normalization and $a_2$ is parameter and fixed to be 
1.4.
Figure \ref{fig3} shows 
the initial velocity distribution in the same case as fig.~\ref{fig2}. 
The flow in longitudinal direction is similar to 
Bjorken's scaling solution in the central region.

\subsection{Calculated Results}
\begin{figure}
\begin{minipage}{0.47\linewidth}
\includegraphics[width= \linewidth]{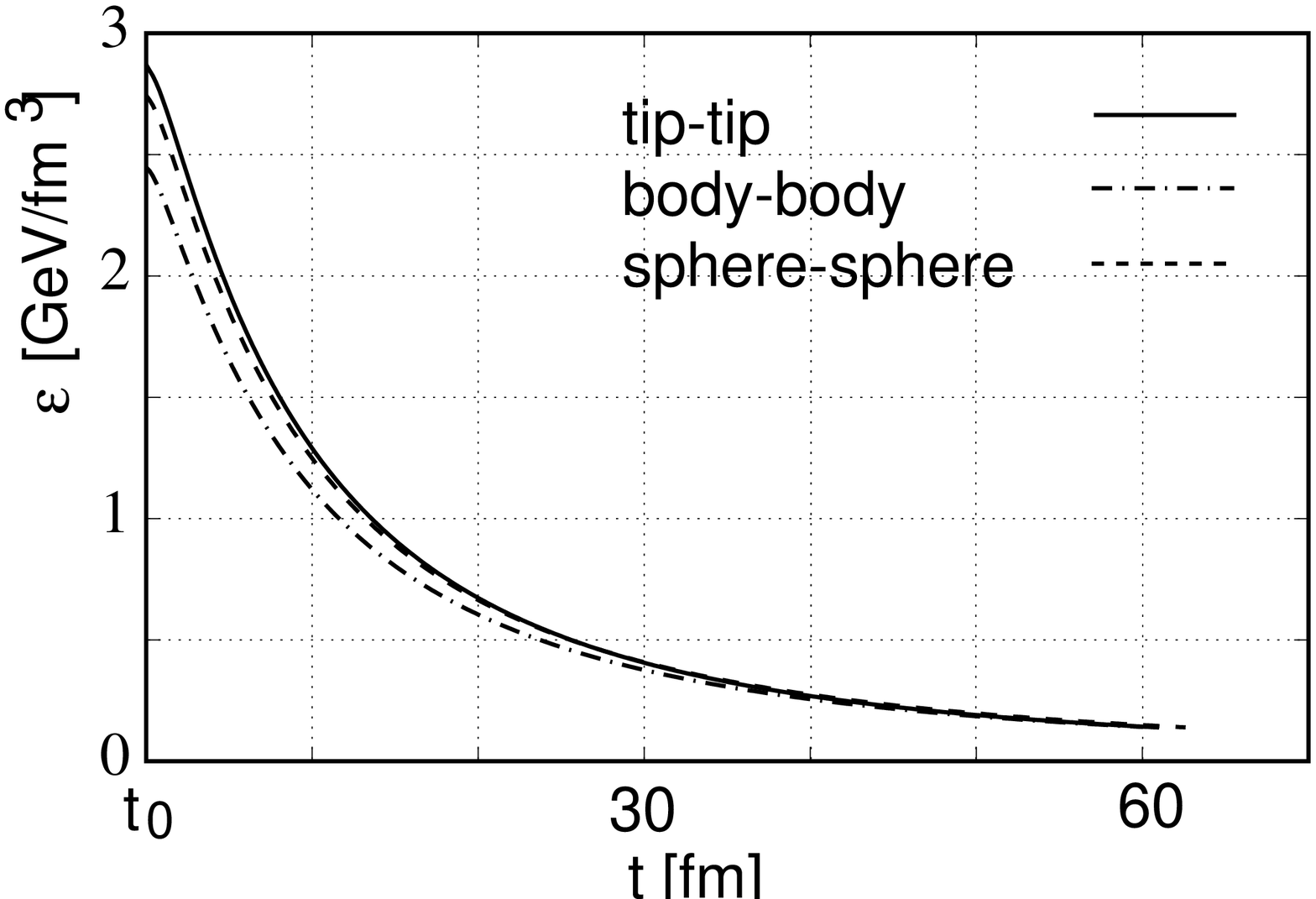}
\caption{The time evolution of energy density of central region 
in U+U 20 AGeV with $b=0$ fm.
The solid line, the dot-dashed line, and the dashed line stand for 
tip-tip collision, body-body collision, and sphere-sphere collision, 
respectively.} 
\label{fig4}
\end{minipage}
\hspace{0.06\linewidth}
\begin{minipage}{0.47\linewidth}
\includegraphics[width= \linewidth]{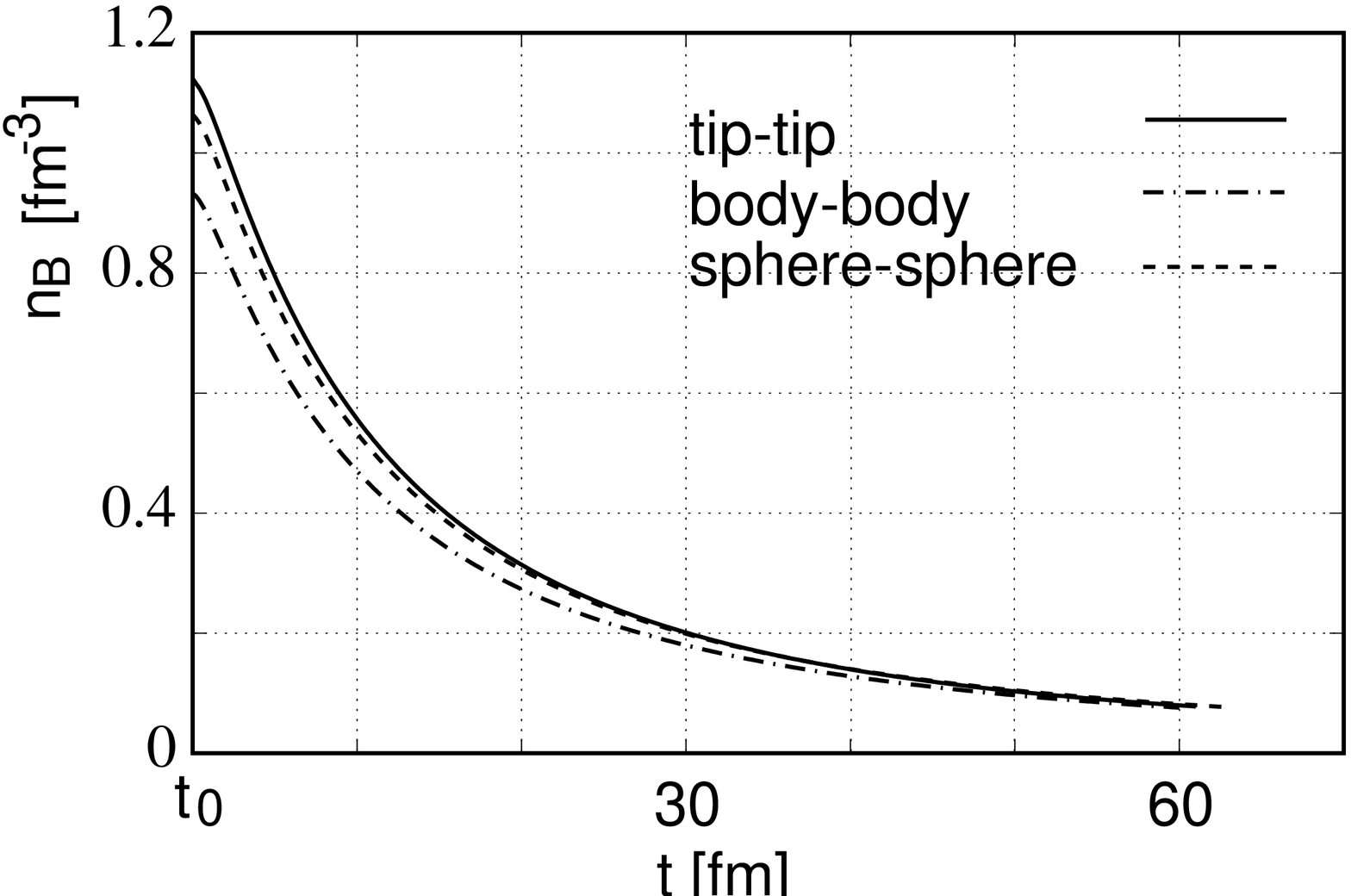}
\caption{The time evolution of baryon number density distribution 
in the same case of  fig.~\ref{fig4}. 
The solid line, the dot-dashed line, and the dashed line stand for 
tip-tip collision, body-body collision, and sphere-sphere collision,  
respectively.}
\label{fig5}
\end{minipage}
\end{figure}
Figures \ref{fig4} and \ref{fig5} indicate the expansion of energy 
density and baryon number density in the central region.
There is a slight difference in the life time in each case. 
The initial energy density and baryon density of body-body 
collision are the smallest among all three types of collisions 
and the difference between tip-tip collision and 
sphere-sphere collision is small.
At initial time the difference of energy density 
among all three types of collision is $\Delta \epsilon \simeq 
10 \sim 20 \%$ and the difference of time is $\Delta t \simeq 10 \sim 
20 \%$ at $\epsilon = 1$ GeV/fm$^3$. However those kinds of difference 
among collision types do not appear at final time.

\begin{figure}
\begin{center}
\begin{minipage}{0.8\linewidth}
\includegraphics[width= \linewidth]{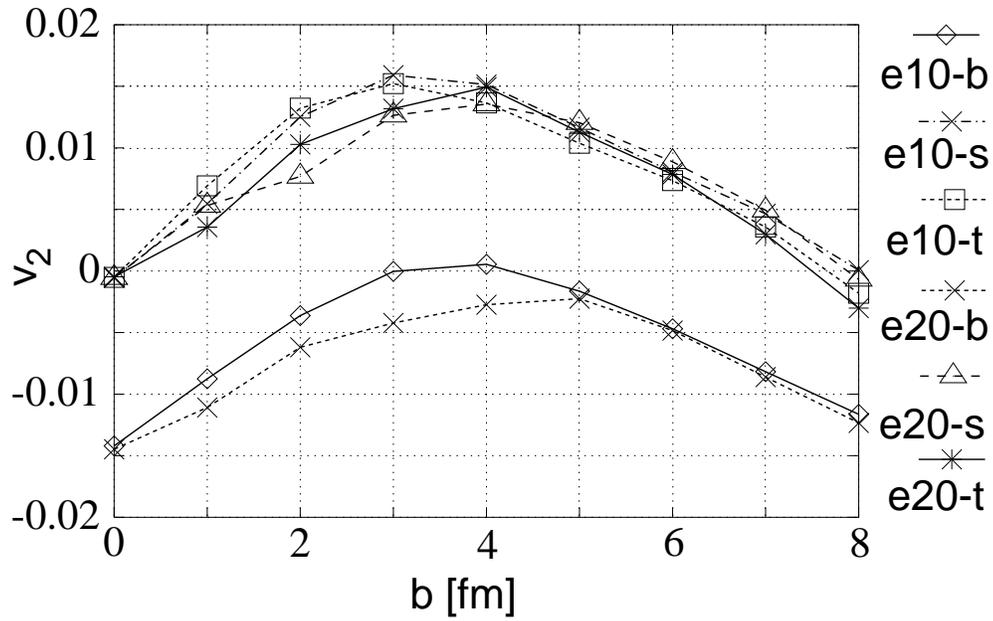}
\caption{The behavior of $v_2$ as a function of the impact 
parameter in all cases. 
In this figure e10-b means that the incident energy is 10 AGeV in 
body-body collision and e10-s means that the incident energy is 
10 AGeV in sphere-sphere collision and e20-t means that the incident 
energy is 20 AGeV in tip-tip collision and so on. }
\label{fig6}
\end{minipage}
\end{center}
\end{figure}
Figure \ref{fig6} shows that behavior of $v_2$ of nucleon 
as a function of the impact parameter for each type of collision 
at 10 and 20 AGeV. 
The elliptic flow parameter, $v_2$, increases with impact 
parameter and reaches 
a peak at $b=3, 4$ fm and decreases in every case.
Furthermore for body-body collision $v_2$ does not vanish at $b=0$. 
Consequently, the effect of deformation is not negligible.  
In order to make this characteristic behavior clear we focus on the 
pressure distribution.

\begin{figure}[ht]
{\Large $b$ = 0 fm}

\vspace{-2.5cm}
\begin{minipage}{.50 \linewidth}
\includegraphics[width= \linewidth]{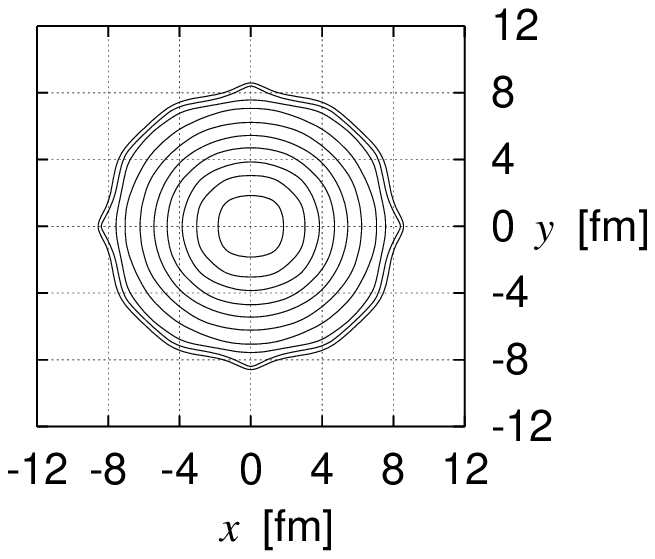}
\end{minipage}
\hspace{-1.5cm}
\begin{minipage}{.50 \linewidth}
\vspace{0.8cm}
\includegraphics[width= \linewidth]{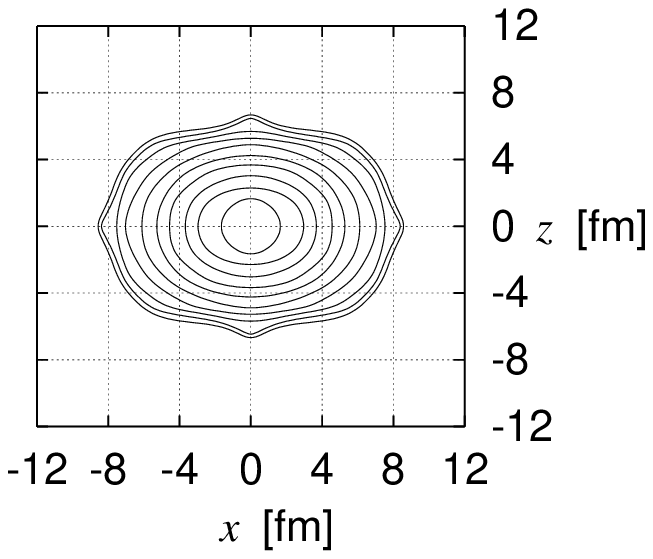}
\begin{center}
\end{center}
\end{minipage}
\vspace{-1.6cm}

{\Large $b$ = 3 fm}

\vspace{-1.7cm}
\begin{minipage}{.50 \linewidth}
\includegraphics[width=\linewidth]{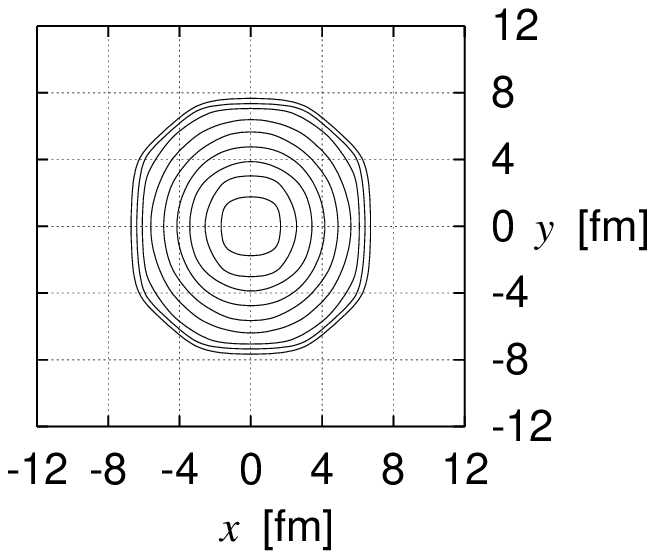}
\begin{center}
\end{center}
\end{minipage}
\hspace{-1.5cm}
\begin{minipage}{.50 \linewidth}
\includegraphics[width=\linewidth]{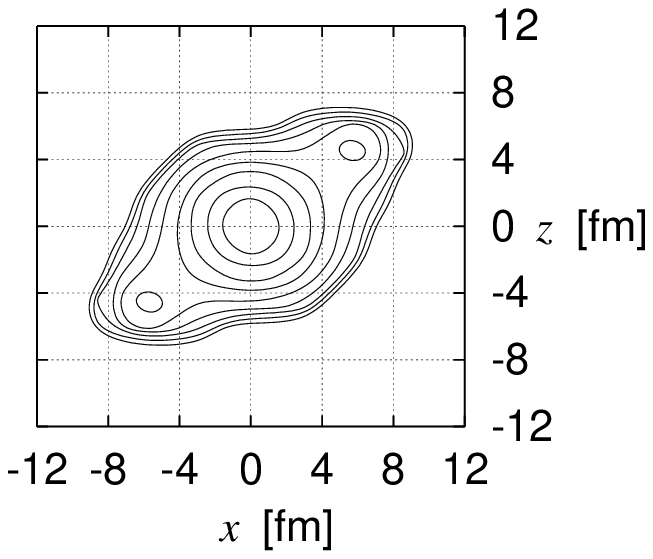}
\begin{center}
\end{center}
\end{minipage}
\vspace{-1.6cm}

{\Large $b$ = 8 fm}

\vspace{-1.7cm}
\begin{minipage}{.50 \linewidth}
\includegraphics[width=\linewidth]{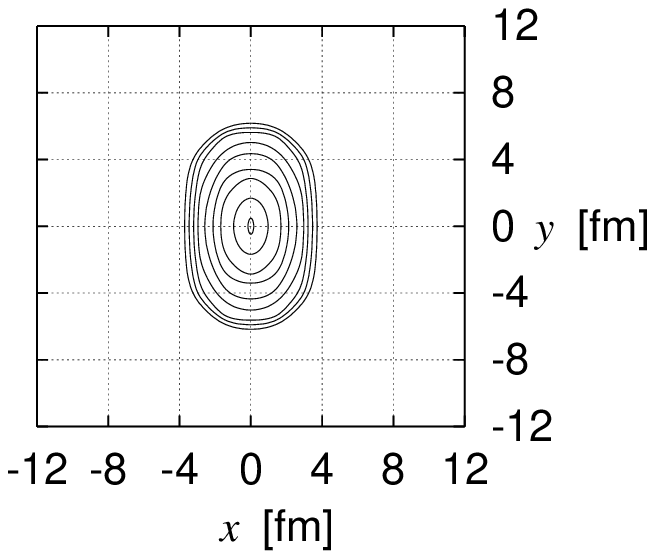}
\begin{center}
\end{center}
\end{minipage}
\hspace{-1.5cm}
\begin{minipage}{.50 \linewidth}
\includegraphics[width=\linewidth]{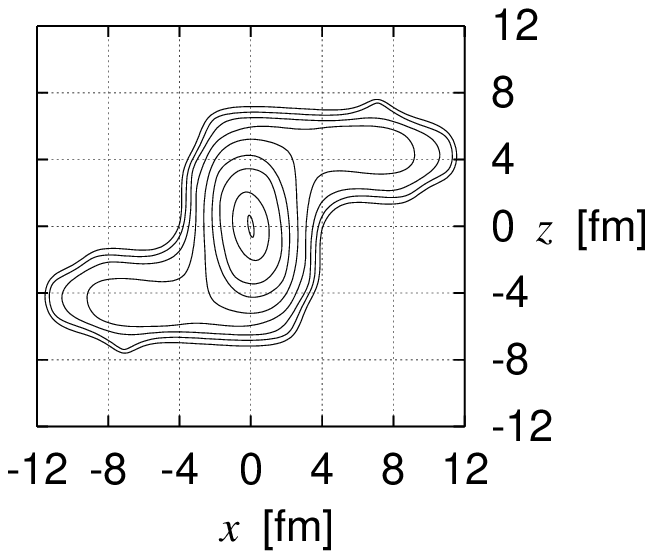}
\begin{center}
\end{center}
\end{minipage}
\vspace{-2cm}
\caption{The impact parameter dependence of the pressure distribution 
in U+U 10 AGeV sphere-sphere collision at $t=20$ fm/$c$. 
The left part of figures displays the
 results at $z \sim 0$ fm and the right part of figures displays those
 at $y \sim 0$ fm. 
The highest values of the pressure
 distribution contour are 0.055 GeV/fm$^3$ for $b=0$ fm, 0.05 GeV/fm$^3$
for $b=3$ and 8 fm,  respectively and the contour lines are drawn 
in steps of $\Delta p =$ 0.005 GeV/fm$^3$.  } 
\label{fig7}
\end{figure}

Figure \ref{fig7} displays the pressure distribution in U+U 10 AGeV 
sphere-sphere collision. 
From the figures at $z \sim 0$ fm we can see that the pressure gradient 
in the $y$ direction increases with the impact parameter. 
Therefore one might consider that $v_2$ increases with impact parameter,
because the velocity of produced particles in $x$ direction is larger than 
one in $y$ direction.
But fig.~\ref{fig6} shows that the value of $v_2$ starts to decrease 
at about $b=3 \sim 4 $ fm. 
Here we focus on the pressure distribution in $z$-plane 
in fig.~\ref{fig7}, and we can see that the effect of spectator increases 
with the impact parameter. Because the spectators block the flow in 
$x$ direction, the growth of flow in $x$ direction is suppressed. 
Consequently the behavior of $v_2$ is determined by both of the 
pressure gradient and the effect of spectator.
 
\begin{figure}[ht]
{\Large tip-tip}

\vspace{-2.5cm}
\begin{minipage}{.50 \linewidth}
\includegraphics[width=\linewidth]{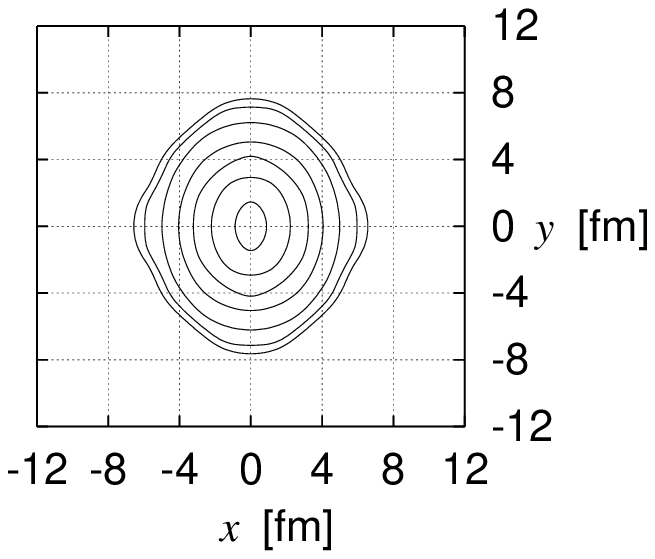}
\end{minipage}
\hspace{-1.8cm}
\begin{minipage}{.50 \linewidth}
\vspace{0.8cm}
\includegraphics[width=\linewidth]{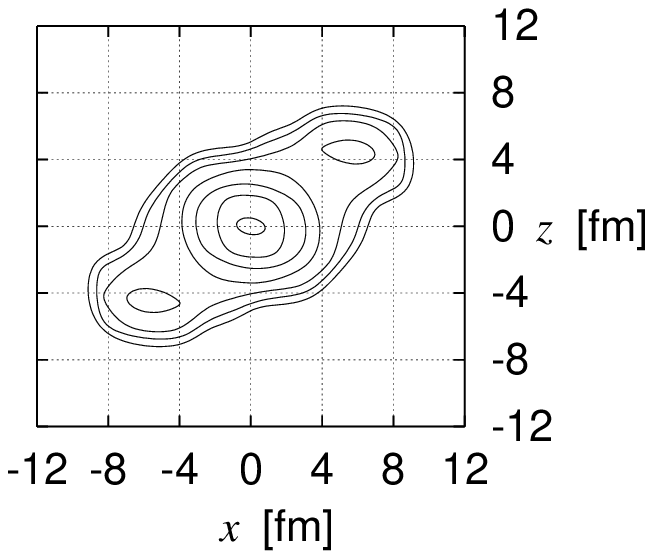}
\begin{center}
\end{center}
\end{minipage}
\vspace{-1.6cm}

{\Large sphere-sphere}

\vspace{-1.8cm}
\begin{minipage}{.50 \linewidth}
\includegraphics[width=\linewidth]{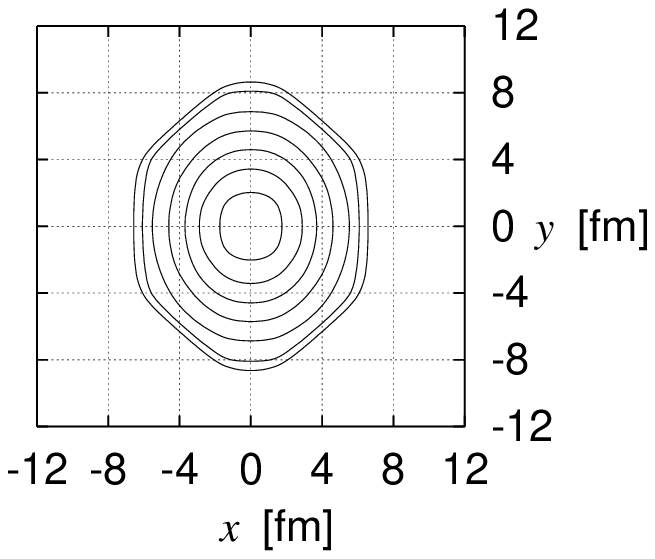}
\begin{center}
\end{center}
\end{minipage}
\hspace{-1.8cm}
\begin{minipage}{.50 \linewidth}
\includegraphics[width=\linewidth]{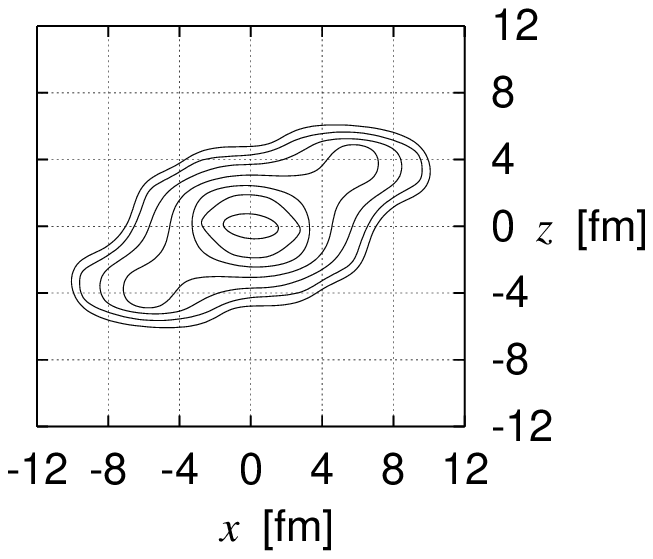}
\begin{center}
\end{center}
\end{minipage}
\vspace{-1.6cm}

{\Large body-body}
\vspace{-1.8cm}

\begin{minipage}{.50 \linewidth}
\includegraphics[width=\linewidth]{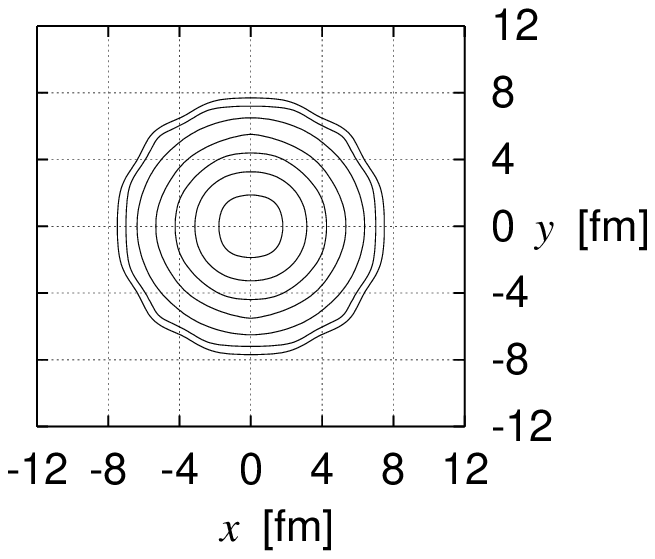}
\begin{center}
\end{center}
\end{minipage}
\hspace{-1.8cm}
\begin{minipage}{.50 \linewidth}
\includegraphics[width=\linewidth]{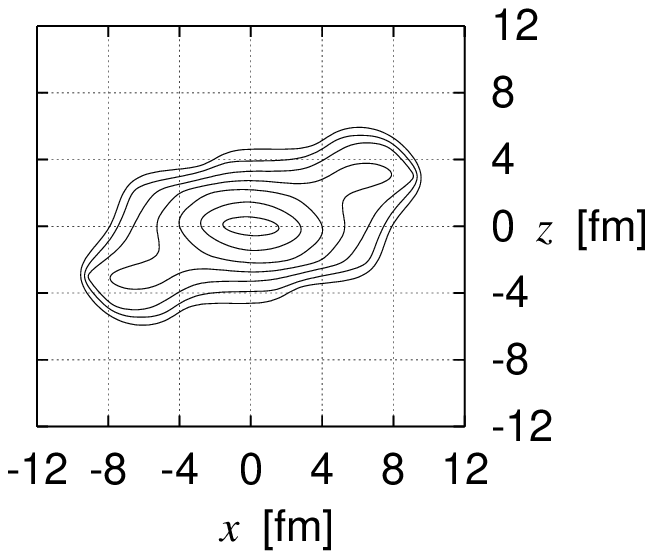}
\begin{center}
\end{center}
\end{minipage}
\vspace{-2cm}
\caption{The pressure distribution in tip-tip, sphere-sphere and 
body-body collision U+U 20 AGeV at $t=20$ fm/$c$. 
The impact parameter is 4 fm in all figures.
The highest value of pressure distribution contour is 0.08 
GeV/fm$^3$ for all figures and the contour lines are drawn in steps 
of $\Delta p = $ 0.01 GeV/fm$^3$. }
\label{fig8}
\end{figure}

Figure \ref{fig8} shows the pressure distribution 
at U+U 20 AGeV collision at $b = 4.0$ fm. 
In $z$-plane there is a slight difference between tip-tip collision and 
sphere-sphere collision. On the other hand, in body-body collision 
the extension of the pressure distribution in $x$ direction 
is larger than in other cases. 
Therefore $v_2$ in body-body collision is less than one in 
sphere-sphere and tip-tip collision at any impact parameter.

\begin{figure}[ht]
{\Large $E$ = 10 AGeV}

\vspace{-2.0cm}
\begin{minipage}{.50 \linewidth}
\includegraphics[width=\linewidth]{2000-z00-10-8-3.eps}
\end{minipage}
\hspace{-1.8cm}
\begin{minipage}{.50 \linewidth}
\vspace{0.8cm}
\includegraphics[width= \linewidth]{2000-y00-10-8-3.eps}
\begin{center}
\end{center}
\end{minipage}

\vspace{-0.5cm}

{\Large $E$ = 20 AGeV}

\vspace{-1.8cm}
\begin{minipage}{.50 \linewidth}
\includegraphics[width=\linewidth]{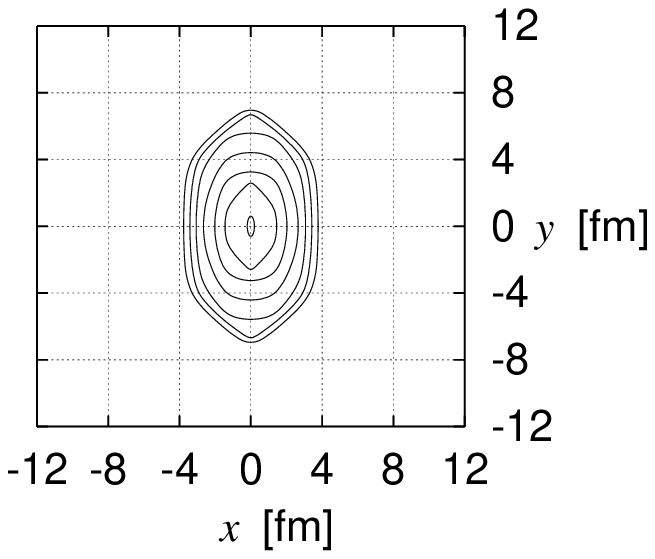}
\begin{center}
\end{center}
\end{minipage}
\hspace{-1.8cm}
\begin{minipage}{.50 \linewidth}
\includegraphics[width=\linewidth]{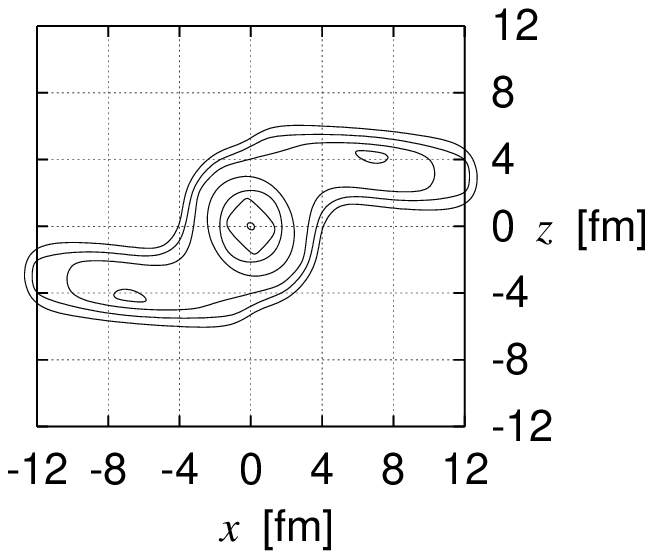}
\begin{center}
\end{center}
\end{minipage}
\vspace{-2cm}
\caption{The incident energy dependence of 
pressure distribution in U+U 10.0 AGeV and U+U 20.0 AGeV at $t=20$ fm/$c$.
The highest value of the pressure distribution is 0.05 
GeV/fm$^3$ in 10.0 AGeV collision  and  0.07 GeV/fm$^3$ in 
20.0 AGeV collision.
The contour lines are drawn in steps $\Delta p = $ 0.005 GeV/fm$^3$ in 10.0 
AGeV collision and in steps $\Delta p =$ 0.01 GeV/fm$^3$ in 
20.0 AGeV collision.}
\label{fig9}
\end{figure}

Figure \ref{fig9} shows the pressure distributions at U+U 10 AGeV and 
U+U 20 AGeV. 
Since there is a slight difference between U+U 10 AGeV and U+U 20 AGeV 
in the $z$-plane, the growth of $v_2$ in U+U 10 AGeV collision is similar
to U+U 20 AGeV.  In the $y$-plane the effect of spectator in 
U+U 20 AGeV is smaller than U+U 10 AGeV. 
Because the effect of Lorentz contraction becomes large in the large 
incident energy, the spectator becomes thin and the influence of it becomes 
small. Consequently the peak of $v_2$ moves to the large impact
parameter at large incident energy.

Here, the value of $v_2$ which is influenced by deformation 
is about 0.015 from fig.~\ref{fig6}. 
According to the analyses of the excitation function of the 
directed flow in ref. \cite{Rischke2, Brachmann},  
the flow becomes slow under phase
transition, because the speed of sound becomes small 
in mixed phase, but the effect of the phase transition on flow is 
expected to be small in the 
AGS energy region.
Therefore the effect of deformation is significant and    
the value of $v_2$ which is obtained in this section is important.
Further detailed analysis on the  
relation between the effect of phase transition and deformation 
on flow may provide a new possible 
experimental probe of the phase transition.  

\newpage
\vspace{-0.3cm}

\vspace{-0.3cm}


\vspace{-0.3cm}

\newpage
\section{Summary}
We present (3+1)-dimensional relativistic hydrodynamical model 
of the Lagrangian hydrodynamics without assuming symmetrical conditions.
Our algorithm is based on the entropy conservation law and the baryon 
number conservation law explicitly. In our algorithm we trace the 
volume elements of fluid along the stream of flux. 
By using our relativistic hydrodynamical model based on the Lagrangian 
hydrodynamics, the path of the each volume element in the phase 
diagram is able to be traced quite easily. 
Therefore we can investigate directly how the phase transition 
takes place and  
affects physical phenomenon in an ultra-relativistic nuclear collision.

Using this model, we have investigated the effect of anisotropic flow in  
deformed uranium collisions.
The behavior of the flow depends on the pressure distribution  
and shadowing. Especially shadowing effect increases with 
the impact parameter.
As for differences in collision types,   
there exists only a slight difference between tip-tip collision and 
sphere-sphere collision.
On the other hand, $v_2$ in body-body collision is
different from sphere-sphere collision, i.e. absolute value of $v_2$ is 
maximum at $b=0$ fm.
In light of the effect of the incident energy on $v_2$, 
the peak of $v_2$ is shifted to 
large impact parameter region, because the shadowing effect decreases 
with increasing the incident energy.
We accordingly conclude that body-body collision is promising for studying
the nuclear equation of states and the property of QCD phase
transition. 
These results are consistent with ref.~\cite{B-A.Li}.      

We shall remark the following things:
In actual experiment, U+U collision is the superposition of different types
of collision like tip-tip, body-body and so on.
Since the contribution of body-body collision is not negligible, 
$v_2$ does not equal to zero at $b=0$ fm.
Furthermore, the shadowing has a large effect on $v_2$.

In this paper we have applied our relativistic hydrodynamical model to 
the anisotropic flow. 
Many kinds of applications of our relativistic hydrodynamical model 
can be considered.  
Using our hydrodynamical model, we can analyze directly 
the phenomena which are sensitive to the phase transition. 
For example, we can argue how the phase transition has effect to 
the minimum of the excitation function of the directed flow, 
discussing the trajectory of volume element of fluid in 
$T$-$\mu$ plane.
Whether ``nutcracker'' phenomena can be observed in our 
Lagrangian hydrodynamical model is also interesting problem. 
Hanbury Brown-Twiss effect (HBT) and the 
influence of anisotropic flow to HBT can also be investigated. 

{\bf Acknowledgment}
The authors would like to thank prof. Osamu Miyamura for his encouragement
to pursue this project and fruitful discussions. 
They thank Dr.~Nobuo Sasaki for providing URASiMA and useful 
comments.
They are grateful to prof. Atsushi Nakamura for reading 
the manuscript carefully.
Calculations have been done at the HSP system 
of the Institute for Nonlinear Sciences and Applied Mathematics, 
Hiroshima University. This work is supported by the Grant in Aide 
for Scientific Research of the Ministry of Education and Culture in 
Japan. (No.~11440080)

\end{document}